\documentclass[fleqn,usenatbib]{mnras}

\usepackage{newtxtext,newtxmath}
\usepackage{multirow}
\usepackage{booktabs}
\usepackage{threeparttable}
\usepackage{siunitx}
\usepackage{caption}

\usepackage[T1]{fontenc}

\DeclareRobustCommand{\VAN}[3]{#2}
\let\VANthebibliography\thebibliography
\def\thebibliography{\DeclareRobustCommand{\VAN}[3]{##3}\VANthebibliography}

\usepackage{graphicx}	
\usepackage{amsmath}	

\title[Short title, max. 45 characters]{MNRAS \LaTeXe\ template -- title goes here}

\author[K. T. Smith et al.]{
Keith T. Smith,$^{1}$\thanks{E-mail: mn@ras.org.uk (KTS)}
A. N. Other,$^{2}$
Third Author$^{2,3}$
and Fourth Author$^{3}$
\\
$^{1}$Royal Astronomical Society, Burlington House, Piccadilly, London W1J 0BQ, UK\\
$^{2}$Department, Institution, Street Address, City Postal Code, Country\\
$^{3}$Another Department, Different Institution, Street Address, City Postal Code, Country
}

\date{Accepted XXX. Received YYY; in original form ZZZ}

\pubyear{2023}

\title{HAYATE: Photometric redshift estimation by hybridising machine learning with template fitting}

\author[S. Tanigawa et al.]{
Shingo Tanigawa,$^{1}$\thanks{E-mail: stanigawa@swin.edu.au}
K. Glazebrook,$^{1}$
C. Jacobs,$^{1}$
I. Labbe,$^{1}$
A. K. Qin$^{2}$
\\
$^{1}$Centre for Astrophysics and Supercomputing, Swinburne University of Technology, PO Box 218, Hawthorn, VIC 3122, Australia\\
$^{2}$School of Software and Electrical Engineering, Swinburne University of Technology, PO Box 218, Hawthorn, VIC 3122, Australia\\
}

\begin{document}
\maketitle

\begin{abstract}
Machine learning photo-z methods, trained directly on spectroscopic redshifts, provide a viable alternative to traditional template fitting methods but may not generalise well on new data that deviates from that in the training set.
In this work, we present a Hybrid Algorithm for WI(Y)de-range photo-z estimation with Artificial neural networks and TEmplate fitting (HAYATE), a novel photo-z method that combines template fitting and data-driven approaches and whose training loss is optimised in terms of both redshift point estimates and probability distributions.
We produce artificial training data from low-redshift galaxy SEDs at $z<1.3$, artificially redshifted up to $z=5$.
We test the model on data from the ZFOURGE surveys,
demonstrating that HAYATE can function as a reliable emulator of EAZY for the broad redshift range beyond the region of sufficient spectroscopic completeness.
The network achieves precise photo-z estimations with smaller errors ($\sigma_{\textrm{NMAD}}$) than EAZY in the initial low-z region ($z<1.3$), while being comparable even in the high-z extrapolated regime ($1.3<z<5$).
Meanwhile, it provides more robust photo-z estimations than EAZY with the lower outlier rate ($\eta_{0.2}\lesssim 1\%$) but runs $\sim100$ times faster than the original template fitting method. We also demonstrate HAYATE offers more reliable redshift PDFs, showing a flatter distribution of Probability Integral Transform scores than EAZY.  The performance is further improved using transfer learning with spec-z samples.
We expect that future large surveys will benefit from our novel methodology applicable to observations over a  wide redshift range.
\end{abstract}

\begin{keywords}
galaxies: distances and redshifts -- methods: data analysis -- techniques: photometric 
\end{keywords}

\label{•} \section{Introduction}

Wide-field imaging surveys are a fundamental driver of astronomical discovery in the fields of galaxy evolution and cosmology.
Galaxy redshifts are a key component in the application of the large-survey data, representing the measurement of galaxy distances.
They are crucial for identifying objects present in the early Universe, tracing the evolution of galaxy properties over cosmic time and constraining cosmological models.

There are two major methods for determining galaxy redshifts: using spectroscopically identified spectral line features (spectroscopic redshifts, hereafter spec-z\textquoteright s), or via multiband photometry \citep[photometric redshifts,][hereafter photo-z\textquoteright s]{Baum62,Butchins81,Connolly+95,Hildebrandt+10}.
Spec-z's are typically much more accurate but more observationally costly than photo-z's; there is a trade-off between the sample size of a dataset and the precision of redshift estimates \citep{Salvato+19}.
In the context of upcoming large surveys, extragalactic astronomy will benefit from photo-z estimation at an unprecedented level as follow-up spectroscopy can never keep pace with present and future large imaging surveys, e.g., the Vera C. Rubin Observatory’s Legacy Survey of Space and Time \citep[LSST;][]{LSSTScienceCollab+09}, the Dark Energy Survey \citep[DES;][]{DarkEnergySurveyCollab+16}, the Nancy Grace Roman Space Telescope \citep[][]{Spergel+15}, the James Webb Space Telescope \citep[JWST;][]{Finkelstein+15}, the Hyper Suprime-Cam Subaru Strategic Program \citep[HSC-SSC;][]{Aihara+18,Aihara+22}, the Euclid mission \citep{EuclidCollab+20} and the Kilo-Degree Survey \citep[KiDS;][]{Hildebrandt+21}.
Thus, efficient and accurate estimation of photo-z\textquoteright s is a topic that has fundamental importance in various fields of research.

There are two main approaches to photo-z estimation.
One is the template fitting method, a kind of model fitting approach \citep[e.g.,][]{Arnouts+99,Benitez00,Bolzonella+00,Feldmann+06,Brammer+08,Eriksen+19}, while the other is the data-driven method of empirical modelling based on spec-z's---machine learning \citep[ML; e.g.,][]{CarrascoKind_Brunner13,Graff+14,Almosallam+1610,Sadeh+16,Cavuoti+1702,Izbicki+17,Graham+18}. 
The main advantage of template fitting is it can be generally applied at any redshift.
It is, however, unable to learn from data to improve performance, which is fundamentally constrained by the template set.
In contrast, the benefit of the data-driven method is generalisation to "unseen data" via learning from the given dataset.
It potentially outperforms template fitting by learning a mapping from photometry to redshift and bypassing potentially unrepresentative templates.
This also reduces the computational demands for photo-z estimation compared to the one-on-one matching between individual objects and templates. However, it typically can not be expected to work outside the redshift range present in the spec-z training set.

Template fitting methods, in which photo-z's are derived from matching the broad- or medium-band photometry of an observed galaxy to pre-defined SED templates, have proven to be very useful.
The template library commonly employed for photo-z study has been updated over the past few decades, exploiting observed \citep[e.g.,][]{Bolzonella+00,Ilbert+06,Ilbert+09,Salvato+09} and synthetic \citep[e.g.,][]{Carnall+18,Battisti+19,Boquien+19,Bellstedt+20} galaxy SEDs.
With this method we can estimate photo-z's for any region of colour space at any redshift.
However, the photo-z estimation with this technique still relies on a limited set of pre-defined templates (which may be more or less representative of the observed galaxy population) as well as the fitting algorithm.
The template fitting method is likewise often computationally intensive and inappropriate for the ongoing and future large survey projects, which would require feasible solutions for analysing unprecedentedly large datasets in peta-scale regimes depending on the science cases.

ML techniques employ an algorithmic model for learning from a given dataset to capture its underlying patterns and then utilise the learned model to make predictions on new data.
They are able to learn from large volumes of data and automatically capture inherent patterns therein that may not be apparent to humans.
In the context of photo-z prediction, this represents a promising route to estimate redshifts from an unprecedentedly huge dataset composed of multi-band photometric data associated with spec-z information.

Different ML algorithms have been utilised in previous works on photo-z estimation.
\cite{CarrascoKind_Brunner13} introduced a photo-z method based on prediction trees and random forest (RF) techniques \citep{Breiman_Schapire01}.
The Multi Layer Perceptron with Quasi Newton Algorithm \citep[MLPQNA;][]{Brescia+13,Brescia+14} contributed to many photo-z works as an excellent demonstration of feed-forward neural networks.
\cite{Sadeh+16} applied multiple ML methods to their model that utilises artificial neural networks (ANNs) and boosted decision trees, while \cite{Jones_Singal17} presented a Support Vector Machine (SVM) classification algorithm for photo-z estimation.
These photo-z based ML methods are generally trained to learn the complex relationship between photometry and distance of observed galaxies.
Most of them have been actually tested on the publicly available data from the PHoto-z Accuracy Testing (PHAT) program \citep{Hildebrandt+10,Cavuoti+12}, performing comparably in terms of photo-z accuracy.

ANNs have been one of the most popular ML algorithms used in photo-z study, which are inspired by the biological neural networks of the human brain \citep{McCulloch_Pitts43,Hopfield82}.
They can theoretically approximate any complex function based on the Universal Approximation Theorem \citep{Cybenko89, Hornik91}, allowing a model to map nonlinear relationships between photometry and redshift.
In particular major advances have been produced exploiting the flexibility of fully connected neural networks (FCNNs), in which each neuron in one layer is connected to all neurons in the next layer.

A major stumbling block for photo-z based ML approaches is incompleteness in spectroscopic training samples commonly used as the ground truth redshift. This limitation could prevent a trained model from functioning as intended, i.e. generalising robustly to new examples outside the training set.
In particular, spec-z catalogues used for training are typically biased towards the bright part of the magnitude parameter space and are incomplete for high-z objects as well.
This also explains why photo-z estimations at high redshifts still rely on existing template fitting methods rather than ML techniques, although they are more common at $z\lesssim1$.
Moreover, training-based methods do not generally allow for reliable extrapolation beyond a known range of data that can be well represented by the training data.
The target redshift range for ML is therefore limited to low-z regions of sufficient spectroscopic completeness with higher success rate in obtaining accurate redshifts for brighter objects.

Furthermore, both template- and ML-based photo-z codes generally fall short in producing valid probability density functions (PDFs) of redshift, which fully characterise the results of photo-z estimation \citep{Schmidt+20}.
Per-galaxy photo-z PDFs have been commonly applied to estimate the ensemble redshift distribution $N(z)$ of a sample of galaxies, an estimator critical to cosmological parameter constraints from weak gravitational lensing analysis \citep[e.g.,][]{Mandelbaum+08,Sheldon+12,Bonnett+16,Hildebrandt+17}.
\cite{Schmidt+20} demonstrated the vulnerability of each single model to a specific flaw in the population of output PDFs in spite of the precise photo-z point estimates.
We still lack a model that can produce well-calibrated redshift PDFs and that can be readily adapted to new studies of galaxy evolution and cosmology.

\citet{wolfBayesianPhotometricRedshifts2009} proposed an example solution for producing accurate redshift distributions from stacked PDFs, although addressing not typical galaxies but specifically quasars under certain conditions. Combining $\chi^2$ template fits and empirical approaches likely preserve both benefits in one framework; empirical training sets can complement unreliable PDFs generated with the $\chi^2$ technique based on imperfect templates if matching the distribution and calibration of query samples. This, however, essentially requires an appropriate treatment of error scale used for smoothing the appearance of samples in feature space and controlling the width of derived PDFs.

Traditional ML approaches have generally delivered better performance than template-based methods within the range of training spec-z coverage \citep{Newman_Gruen22}.
The trade-off between the strengths of ML and template fitting inspires the hybridisation of their distinctive advantages.
Training the model on simulated photometry is one strategy to overcome the challenges of assembling a complete, reliable and unbiased training sample of sufficient size.
Artificial SED samples are often generated using a stellar population synthesis (SPS) code with arbitrary selection of free parameters \citep[e.g.,][]{Eriksen+20, Ramachandra+22}.
\cite{Zhou+21} applied a set of best-fit SEDs for the COSMOS catalogue using the template fitting code \texttt{LePhare}, produced based on typical SPS spectra \citep{Bruzual_Charlot93, Bruzual_Charlot03}.
A complete training set of simulated galaxies should compensate for the sparse sampling of spec-z data
allowing for interpolation between spectroscopically observed objects and even extrapolation to the faintest ones \citep{Newman_Gruen22}.
The fidelity of the mock training samples is still liable to many stellar evolution uncertainties that have long plagued SPS models \citep{Conroy13}.
Constructing such an ideal SED dataset requires further improvements to SPS models and to our knowledge of the underlying galaxy population.

Alternatively, the template fitting code EAZY \citep{Brammer+08} provides more flexible galaxy SEDs, which fits a linear combination of basic spectral templates to the observed photometry on-the-fly.
They developed a minimal template set of synthetic SEDs representing the ``principal components'', following the template-optimisation routines introduced by \cite{Blanton_Roweis07}.
The template set is calibrated with semi-analytical models rather than biased spectroscopic samples, which are complete to very faint magnitudes, along with a template error function to account for wavelength-dependent template mismatch.
The applicability of EAZY to diverse redshift coverage has been demonstrated with a plethora of photometric catalogues \citep[e.g.,][]
{Treistere+09,Wuyts+0905,Cardamone+10,Muzzin+13,Nanayakkara+16,Straatman+16,Strait+21}.
In particular, the reliability of EAZY photo-z's was thoroughly assessed with comprehensive photometric samples presented by \citet[hereafter S16]{Straatman+16}, which include medium-bandwidth filters from the \texttt{FourStar} galaxy evolution (ZFOURGE) surveys.

In this work, we present a novel hybrid photo-z method that combines template fitting and data-driven approaches to exploit the best aspects of both.
Our photo-z network is trained with mock photometric data generated based on the ensemble of template SEDs provided by EAZY.
This is particularly motivated by exploiting knowledge of galaxy SEDs at low-z, where template fitting is assumed to be reliable, and applying their rest-frame SEDs to a higher redshift range.
The full training set of mock SEDs is thus generated by redshifting best-fit SEDs derived with EAZY for the S16 photometric catalogue objects of $z\lesssim 1$, whose simulated redshifts are distributed in a broader range up to $z=5$.
We develop photo-z convolutional neural networks \citep[CNNs;][]{Lecun+98,Lecun+04} optimised to simultaneously produce both a well-calibrated set of redshift PDFs and accurate point estimates.
The trained model is tested with updated S16 spectroscopic samples, whose performance is evaluated based on photo-z metrics commonly used for measuring the quality of both output PDFs and the corresponding point estimates.

Our ML strategy benefits from recent advances in the field of domain adaptation \citep{Csurka17,Wang_Deng18,Wilson_Cook20}, which allows a model to learn domain-invariant features shared between discrepant data distributions.
The simulation-based ML model here is trained with synthetic data, which can be further advanced by transfer learning \citep{Pan_Yang10}, where a model pre-trained on one task is re-purposed on another related task.
Pre-training the feature extraction layers on a large external dataset then fine-tuning on a smaller training set alleviates overfitting compared to simply training from scratch on the small dataset. 
We can thus fine-tune the simulation-based photo-z network with a limited amount of spectroscopic data by re-training the last layers on real datasets with spec-z information \citep{Eriksen+20}.
This optimisation scheme in principle aids in correcting the gap between mock and observed training samples.

Our novel approach is to "extrapolate" training methods outside their initial redshift ranges from the viewpoint of the original template fits.
Training with domain adaptation can be performed on high-z simulated data by capturing a realistic range of galaxy SED properties determined from reliable low-z data.
In place of spectroscopic data we leverage the demonstrated accuracy of template fitting, overcoming the traditional redshift limitation of ML photo-z codes. 
In essence, the CNN-based hybrid model is thus designed to function as an efficient emulator of EAZY.
The interpolative nature of supervised ML approaches could even infer photo-z point estimates more precisely and robustly than those provided by the original template-based method.
Incorporating the flavour of template fitting into the ML framework potentially improves the quality of photo-z PDFs as well.
Ultimately, we aim to improve photo-z estimation for JWST photometry, which will have coverage at redder wavelengths than previously available.

This paper is organised as follows. In \S \ref{sec:catalog_data}, we present the photometric catalogues used in this work.
In \S \ref{sec:Training set of mock photometric data}, we detail our method for producing mock photometric data (with a noise model) via simulations.
\S \ref{sec:ML-photo-z-model} describes the development of our ML photo-z networks and the framework for evaluating their performance.
\S \ref{sec:results} presents results on testing different photo-z models on the ZFOURGE catalogue data and comparing their performance in photo-z and PDF metrics commonly used for major photo-z studies.
In \S \ref{sec:Discussion} we discuss some of the issues raised by the work. Finally, in \ref{sec:Conclusion} we summarise the work and discuss future prospects.
Throughout this paper, we assume a $\Lambda \rm CDM$ cosmology with $\Omega_{M}=0.3$, $\Omega_{\Lambda}=0.7$ and $H_{0}=70\mathrm{km\:s^{-1}\:Mpc^{-1}}$.

\section{Catalogue data}

\label{sec:catalog_data}

This work introduces a hybrid photo-z based ML method that benefits from the template fitting algorithm of EAZY, aimed at deriving photo-z PDFs of galaxies extracted from the ZFOURGE photometric catalogues \citep[][]{Straatman+16}.
ZFOURGE data products comprise 45 nights of observations with the \texttt{FourStar} instrument \citep{Persson+13} on the 6.5 m Magellan Baade Telescope at Las Campanas in Chile.
It observed three survey fields including CDFS \citep{Giacconi+02}, COSMOS \citep{Scoville+07} and UDS \citep{Lawrence+07} with five near-IR medium-bandwidth filters, $J_{1}$, $J_{2}$, $J_{3}$, $H_{s}$, and $H_{l}$, along with broad-band $K_{s}$.
Pushing to faint magnitude limits of 25--26 AB achieves the mass completeness limit of $\sim10^{8}\:M_{\odot}$ at $z\lesssim1$, also advancing the study of intermediate to high redshift objects.

S16 includes data from publicly available surveys
at 0.3-\SI{8}{\micro\meter}, constructing comprehensive photometric
catalogues, each with a total of 39 (CDFS), 36 (COSMOS) and 27 (UDS)
medium- and broad--band flux measurements. The individual objects
were cross-matched with the compilation of publicly available spec-z
catalogues provided by \cite{Skelton+14} as well as the first data
release from the MOSDEF survey \citep{Kriek+15} and the VIMOS Ultra-Deep
Survey \citep{Tasca+17}. These samples have been used to demonstrate
the benefit of including the \texttt{FourStar} medium bands in the
input for improving the photo-z accuracy with a better sampling of
galaxy SEDs \citep{Straatman+16}.

Throughout, the catalogue data utilised for this work are limited
to objects with a \texttt{use} flag of 1, which represents reliable
data with good photometry and a low likelihood of contamination with stars or blending with another source.
These sources are obtained from regions of the images with sufficiently high S/N. 
We thus construct test catalogue samples with \texttt{use} = 1 and
total $K_{s}$-band magnitude $<26$, providing the galaxy population
that can be used in large statistical studies. Our main target objects
are high-z galaxies of $z\gtrsim1.3$, whose photo-z estimations have
not been well explored by ML methods. We set the lower limit to $1.3$
as that is a typical bound for which spec-z\textquoteright s are incomplete,
since the galaxy optical light is redshifted in to the near infra-red. The model
is nonetheless required to make predictions across the whole redshift
range (including lower $z's$), since we cannot exclusively select
high-z objects \textit{a priori} from real observations. Our spec-z samples
are therefore limited only with an upper bound of 5, which are adopted
as a test set for evaluating the model's performance on the broad
redshift range between $0<z_{\textrm{spec}}<5$.

Additionally, we incorporate ancillary spec-z data from the latest releases of several surveys into our original S16 catalog, with a matching radius of 1\arcsec.
All the catalogues are supplemented by the final data releases from the MOSDEF \citep{Kriek+15} and MOSEL \citep{Tran+20,Gupta+20} surveys.
The fourth data release from the VANDELS surveys \citep{Garilli+21} provides auxiliary spec-z's for CDFS and UDS, while the ZFIRE survey \citep{Nanayakkara+16} for COSMOS.
We only extract reliable data with the best quality flag individually defined for each survey catalogue.

As a further step, two of the authors (KG and IL) visually inspected spectra where the spec-z and EAZY photo-zs differed signficantly.  We removed objects deemed likely misidentifications, providing sample sizes
of 1100 (CDFS), 425 (COSMOS) and 127 (UDS) from the original S16 catalogue.
The size of each supplemented sample ($z > 1.3$) is as follows: 1273 in CDFS, 741 in COSMOS and 314 in UDS, an increase of 173, 316 and 187, respectively.

\section{Training set of mock photometric data}

\label{sec:Training set of mock photometric data}

In this section, we discuss the generation of mock photometric data used for training the ML model.
The entire process is divided into two major parts, both of which are important for creating a training sample that can sufficiently cover the colour space occupied by the test sources.
\S \ref{subsec:Redshifting} describes the method of producing mock SEDs from EAZY best-fits for a limited sample of low-z galaxies in S16.
In \S \ref{subsec:noise_model}, the noise model is introduced to apply realistic errors to simulated photometry, which allows for the construction of reliable mock photometric data.

\subsection{Mock galaxy SEDs}

\label{subsec:Redshifting}

We simulate galaxy SEDs up to $z=5$ by redshifting the 
EAZY best-fit SEDs for low-z objects with $z_{\textrm{EAZY}}<1.3$
in S16. This enables us to produce SEDs of galaxies in the target
redshift range between $1.3<z<5$ purely based on a 
galaxy population at lower redshifts. The selection criteria of the
low-z sources also ensures the generated sample fully covers typical
SED types, since ZFOURGE is very complete to low masses at $z\lesssim1.3$,
where the 80\% mass completeness limit reaches down to $\sim10^{8}-10^{8.5}M_{\odot}$
\citep{Straatman+16}. We thus first extract EAZY best-fits for objects
with $z_{\textrm{EAZY}}<1.3$ that are included in the photometric catalogues
of S16. The total number of selected low-z sources is 17,891.
These empirical SEDs are technically unique, since EAZY fits an ensemble
of nine representative spectral templates to each set of observed
fluxes. The major part of our simulated sample thus consists of typical
SED types empirically obtained from low-z observations but assumed
to be present at much higher redshifts.

We then artificially redshift these pre-defined SEDs from the limited
redshift range of $z_{\textrm{EAZY}}<1.3$ to simulated redshifts 
($z_{\textrm{sim}}$'s) in a much broader range of $0<z_{\textrm{sim}}<5$.
For each mock SED, a set of simulated wavelength and flux density 
per unit wavelength $(\lambda_{\textrm{sim}},F_{\textrm{sim}})$ measurements 
are derived from the
EAZY output $(\lambda_{\textrm{EAZY}},F_{\textrm{EAZY}})$ with the following equations:
\begin{eqnarray}
\lambda_{\textrm{sim}} & = & \lambda_{\textrm{EAZY}}\left[\frac{1+z_{\textrm{sim}}}{1+z_{\textrm{EAZY}}}\right],\\ \label{eq:lambda_sim}
F_{\textrm{sim}}(\lambda_{\textrm{sim}}) & = & F_{\textrm{EAZY}}(\lambda_{\textrm{EAZY}})\left[\frac{D_{\textrm{EAZY}}}{D_{\textrm{sim}}}\right]^{2}\left[\frac{1+z_{\textrm{EAZY}}}{1+z_{\textrm{sim}}}\right],\label{eq:F_sim}
\end{eqnarray}

where $D_{\textrm{EAZY}}$ and $D_{\textrm{sim}}$ are the luminosity distance for the
observed and simulated galaxies.

The simulated data are generated with a uniform distribution with
respect to $\zeta=\log(1+z)$, which is adopted as our output variable
instead of the simple redshift \citep{Baldry18}. This adapts to the evaluation scheme
commonly used in most photo-z studies, where the redshift estimation
error is defined as $d\zeta=dz/(1+z)$. Using $d\zeta$ as a reasonable
photo-z error is ascribed to different photometric uncertainties for
a given set of broad-band filters, which typically have an approximately
constant resolution of $R=d\lambda_{obs}/\lambda_{obs}\sim const.$,
where $\lambda_{obs}$ is an observed wavelength. $d\zeta$ thus shows
a constant error if an observational error of $dz$ purely scales
with the filter spacing $d\lambda_{obs}$ while $\lambda_{obs}$ with $(1+z)$.

The uniform distribution of simulated $\zeta$'s ensures that the
number density of the training data is constant at any $\zeta$, which
is required for developing a photo-z network whose error estimations
are not biased in the entire redshift range. One of our goals is to
build a model that produces reliable redshift PDFs as well as single-point
estimates, which is implemented by outputting probabilities for 350
$\zeta$ class bins, as described in \S \ref{subsec:Inputs-and-outputs}.
We generate multiple mock SEDs from a given low-z source by randomly drawing
$\zeta$ in each of 35 equally discretised bins, whose resolution is
10 times lower than the output probability vector. 
The sample
size of our mock SEDs consequently results in $\sim600,000$. 

Our knowledge of the underlying galaxy SEDs is exclusively attributable
to objects observed with the FourStar medium-band filters. The high number of filters in these photometric data
ensures the individuality of
each empirical template, which would be otherwise standardised into
a small set of simplified representations. 
This allows us to efficiently generate realistic high-z SEDs even 
in the absence of large amounts of data about the distant universe.
We note that the current framework does not take into consideration
the difference in population between low-z and high-z galaxies due to
their evolution. Handling this issue in a robust manner
is beyond the scope of this paper, but our input fluxes are normalised
to remove magnitude information, as described in \S \ref{subsec:Inputs-and-outputs},
which should alleviate the impact on the model's performance.

\subsection{Photometry simulations with noise application}

\label{subsec:noise_model}

The photometry for the mock SEDs is simulated using a transmission
curve for each filter adopted in S16, producing a noiseless flux
per unit wavelength $\bar{F}_{i}$ for the band $i$. Establishing
a realistic photometric sample then requires artificially applying
an observational error to each noiseless flux. The fundamental concept
of our fiducial noise model (which we call `empirical') is to introduce actual observational
noise for one test source $t$ into simulated photometry of each mock
SED.

We explore the most appropriate noise realisation for a given simulated SED in
comparison with the observed data.
This requires a measure of similarity in SED shape $S_{t}$ between
noiseless simulations $\bar{F}_{i}$ and noised observations
$(\tilde{F}_{i,t},\tilde{E}_{i,t})$, where $(\tilde{F}_{i,t},\tilde{E}_{i,t})$
is a set of flux and error observed for the band $i$ from the source $t$. An approximate
SED shape is captured by normalising all the fluxes and errors of
each object by its own $K_{s}$-band photometric measurement. Each
pair of simulated and catalogue sources are then compared based on
these normalised photometric data, $\bar{f}_{i}$ and $(\tilde{f}_{i,t},\tilde{e}_{i,t})$ 
(here we denote normalised data with lower case).

For each mock galaxy, the similarity between $\bar{f}_{i}$
and $(\tilde{f}_{i,t},\tilde{e}_{i,t})$ is measured by assuming
each simulated flux $f_{i,t}$ follows a Gaussian distribution given
a standard deviation $\tilde{e}_{i,t}$. EAZY also adapts to template
mismatch with a rest-frame template error function $\sigma_{te}(\lambda)$.
The total flux uncertainty $\delta f_{i,t}$ is given by
\begin{equation}
\delta f_{i,t}=\sqrt{\left.\tilde{e}_{i,t}\right.^{2}+\left[\bar{f}_{i}\sigma_{te}(\lambda_{i,\mathrm{rest}})\right]^{2}},\label{eq:delta_f}
\end{equation}
where $\lambda_{i,\mathrm{rest}}$ is the rest-frame central wavelength of
the filter $i$, expressed with the observed wavelength $\lambda_{i}$
as $\lambda_{i,\mathrm{rest}}=\lambda_{i}/(1+z_{\textrm{sim}})$.

We thus assume $f_{i,t}\sim N\left(\bar{f}_{i},\left.\delta f_{i,t}\right.^{2}\right)$
to estimate a probability $p_{i,t}$ that the observed $\tilde{f}_{i,t}$
is realised, given by
\begin{equation}
p_{i,t}=\frac{1}{\sqrt{2\pi}\delta f_{i,t}}\exp\left[-\frac{1}{2}\left(\frac{\tilde{f}_{i,t}-\bar{f}_{i}}{\delta f_{i,t}}\right)^{2}\right].\label{eq:p_ij}
\end{equation}
The product of fluxes across each band then measures the stochastic similarity
of the mock galaxy to the catalogue source $t$:
\begin{equation}
P_{t}=\prod_{i}^{n_{t}}p_{i,t},\label{eq:Pi}
\end{equation}
where $i$ covers $n_{t}$ broad- and medium-band filters adopted
in S16 which do not contain missing values. The similarity measure
$S_{t}$ consequently needs to be defined in a form that should be
generally applicable to comparing any pairs, since the effective number
of filters $n_{t}$ is not fixed for all the catalogue sources, dependent
on $t$. One reasonable measurement is given by
\begin{equation}
S_{t}=(P_{t})^{1/n_{t}},\label{eq:Si}
\end{equation}
which can function as a probability of realisation for an object
$t$.

We additionally adopt a magnitude prior $p(z|m)$ following \cite{Straatman+16}
for computing a probability of drawing a test source $t$, expressed
as
\begin{equation}
P(z_{\textrm{sim}},t)=S_{t}p(z_{\textrm{sim}}|m_{t}),\label{eq:p_zsim_mt}
\end{equation}
where $m_{t}$ is the $K_{s}$-band apparent magnitude. One catalogue
object is randomly picked with a probability $P(z_{\textrm{sim}},t)$, whose
errors $\{\tilde{e}_{i,t}\}_{i}$ are applied to each simulated SED
including its missing values. The noised flux $F_{i,t}$ is then obtained
by denormalising $f_{i,t}\sim N(\bar{f}_{i},\tilde{e}_{i,t}^{2})$.

\begin{figure*}
\begin{centering}
\includegraphics[width=0.8\textwidth]{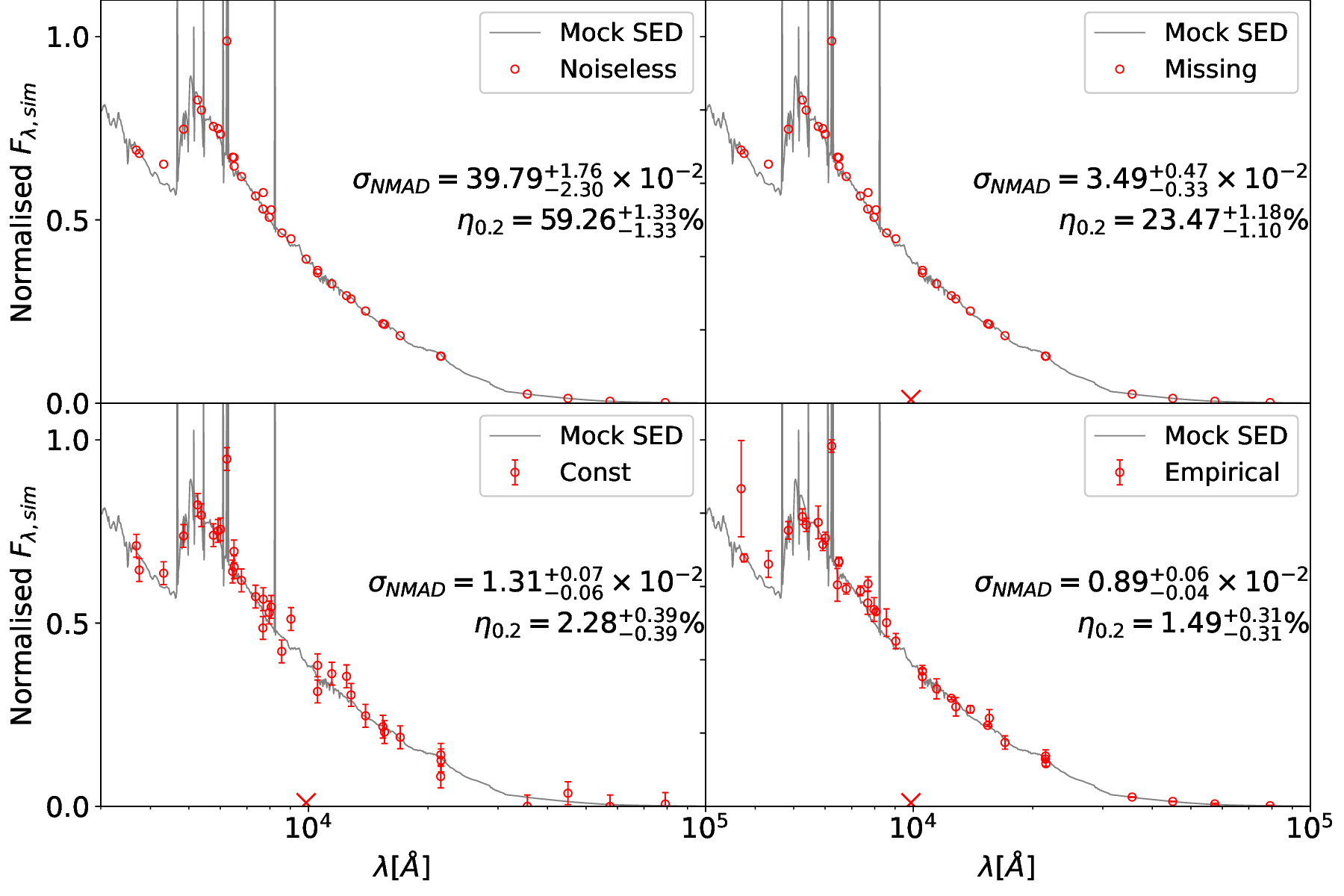}
\par\end{centering}
\caption{Simulated photometry with different noise models for the same mock
SED. All the simulated fluxes are shown by the red circles with error
bars, while the red crosses represent missing data. The top panels
present purely integrated photometry without artificial noise, but
without (left) and with (right) some missing values included based
on a randomly picked catalogue source. These flux points are drawn
from the Gaussian distributions with a constant variance over all
wavelengths in the bottom left panel. The bottom right panel exhibits
the artificial noise generated from the photometric data of a catalogue
source with a similar SED shape. \label{fig:noise_model}}
\end{figure*}

We also establish simpler noise models to explore the benefit of our empirical one:
\begin{enumerate}
\item[(i) Noiseless:] all the noiseless simulated fluxes are fed to the photo-z network as inputs, given by $F_{i}=\bar{F}_{i}$.
\item[(ii) Missing:]  for each mock SED, we randomly draw one test source from the spec-z catalogue whose missing values for some band filters are directly incorporated into the simulated photometric data.
\item[(iii) Const:]  photometry for each mock SED is performed with a constant noise $E_{{cnt}}$ over the entire wavelength range. $E_{cnt}$ is obtained by assuming an arbitrarily selected S/N for $K_{s}$-band photometry, where S/N is a random variable ranging between 3 and 30. Each noiseless flux point then varies following a Gaussian distribution with $F_{i}\sim N\left(\bar{F}_{i},E_{cnt}^{2}\right)$, which also reflects the missing values in the same method as the Missing model (ii).
\item[(iv) Empirical:]  our fiducial model.
\end{enumerate}
Fig. \ref{fig:noise_model} shows the simulated photometry for an
example mock SED, whose noised fluxes are generated with the four
different noise models. The Missing model (ii) drops one flux value
as missing, which is represented by the red cross, while the Const
model (iii) further adds constant errors to the remaining fluxes.
More realistic photometry can be simulated with the Empirical model
(iv), where the empirical noise is applied to the noiseless fluxes
which is extracted from the test sample.

We then train the CNN models, whose architecture is introduced in
\S \ref{subsec:Architecture-of-HAYATE}, on the different simulated
datasets for CDFS, each generated with one of the four noise models.
Testing them on the same spec-z catalogue sample allows us to explore
the most effective noise model. The performance of each CNN is evaluated
with the accuracy $\sigma_{\textrm{NMAD}}$ and the outlier rate $\eta_{0.2}$
of photo-z point estimates, as described in \S \ref{subsec:Indicators-for-photo-z-estimates}.
Fig. \ref{fig:noise_model} presents the results, revealing the
Noiseless model (i) causes a catastrophic failure in photo-z estimations
since the training sample does not contain any errors and missing
values in photometric measurements. This can be improved by incorporating
missing values into the training set which reflect those of the test
sample. The Missing model (ii) achieves much better results of $\sigma_{\textrm{NMAD}}\sim0.03$
and $\eta_{0.2}\sim20\%$ than those of the Noiseless model (i) with
$\sigma_{\textrm{NMAD}}\sim0.4$ and $\eta_{0.2}\sim60\%$.

The Const model (iii) shows further improvements by applying simple
artificial noise to the noiseless fluxes, reducing $\sigma_{\textrm{NMAD}}$
and $\eta_{0.2}$ to $\sim0.013$ and $\sim2.3\%$. Significantly
better scores can be obtained as well by training the model on more
realistic mock data generated with the Empirical model (iv), which
result in $\sigma_{\textrm{NMAD}}\sim0.009$ and $\eta_{0.2}\sim1.5\%$. These
results indicate the empirical noise application shows the smallest
disparity between simulations and observations.
We therefore conclude that the the Empirical model (iv) 
can produce mock photometric data which best represents the test catalogue samples. The empirical treatment of noise in the training set further improves the precision of PDFs derived for the query set, which can translate into matching the error scales of the distinct samples \citep{wolfBayesianPhotometricRedshifts2009}. Effectively, the combination of our chosen noise model, our loss function, and the nonlinearity of the neural networks may allow the model to treat the error scale as a parameter and optimise it such that the smoothing scale of the combined error more effectively matches that of our target data."

We randomly generate five realisations of empirical noise based on the
same mock SED sample for each field. This provides stochastically
different photometric samples, each constructed by matching the given
simulated galaxies with randomly selected catalogue data following the
relative probability $P(z_{\textrm{sim}},t)$. They are independently used
for training different networks, whose predictions are subsequently
combined with the ensemble learning method, as discussed in \S \ref{subsec:Ensemble-learning}.
We note that missing values present in the test catalogue samples
are incorporated into the photometry simulation. This
allows our training set to intrinsically contain information on the
corresponding missing data, which does not require imputing missing
values entailed by the test set for evaluating the model's performance
on real data.

\section{ML photo-z model}

\label{sec:ML-photo-z-model}

We can assess the performance of photo-z networks on the S16 test catalogue
by first training them with the mock data. \S \ref{subsec:Inputs-and-outputs}
describes the input and output, which are designed for yielding redshift
PDFs from normalised photometric data. In \S \ref{subsec:baseline_model}
and \S \ref{subsec:Architecture-of-HAYATE}, the architectures of
two different photo-z networks are introduced: a fully connected neural network (FCNN) and a CNN-based model HAYATE. \S \ref{subsec:Indicators-for-photo-z-estimates}
discusses commonly-used evaluation metrics for photo-z point estimates
and their PDFs. \S \ref{subsec:Training-process} describes the fiducial
training configuration for each network, whose lower-level output
PDFs are combined with the ensemble learning method, as discussed
in \S \ref{subsec:Ensemble-learning}. In \S \ref{subsec:Transfer-learning},
we discuss the benefit of transfer learning using spec-z's for further
improvements.

\subsection{Inputs and outputs}

\label{subsec:Inputs-and-outputs}

Our training set contains simulated high-z galaxies which mirror the 
pupulation of low-z ZFOURGE sources; no evolution of the galaxy 
population is accounted for.
We thus remove information on magnitudes from the input, which are critically influenced by the formation and evolution of galaxies and highly correlated with redshift. 
Each galaxy
is consequently identified purely based on its SED shape. Our input
variables are thus primarily flux ratios, which are obtained for each
galaxy by normalising photometric measurements with its total $K_{s}$-band
flux provided by S16. The photometry is a product of
stacked\texttt{FourStar}/$Ks$-band and deep pre-existing $K$-band imaging.
The super-deep image achieves a maximum limiting depth at $5\sigma$
significance of $26.2-26.6$, $25.5$ and $25.7$ mag in CDFS, COSMOS
and UDS, respectively. Using the total $Ks$-band flux as a baseline
therefore ensures the normalised datasets are reliably calibrated. A similar
scheme has been established by \cite{Eriksen+20} with input fluxes
divided with the $i$-band flux. 

Testing a trained model on the spec-z catalogue also requires handling missing values, which are inevitably present in real data.
We adopt a standard approach of imputation by a constant value, replacing all missing values in the normalised input data with -1.
The negative substitute value can exclusively represent a lack of effective data distinguished from the other flux measurements, which should be zero or more.
As depicted in Fig. \ref{fig:noise_model}, our missing data replacement strategy, represented by the Missing model (ii) described in $\S$ \ref{subsec:noise_model}, markedly improves the model's performance compared to other imputation methods. 
We note that each data point can potentially represent no flux measurement as distinct from a missing value, an important distinction when mapping from photometry to redshifts.
Therefore using a zero value is not appropriate as a placeholder for missing data.
We could also employ a more complex method to 
substitute missing values, depending on the individual dataset, such as interpolation/extrapolation and k-nearest neighbours.
As these approaches generate fake (though plausible) values for imputation, they could potentially degrade the precision of estimated photo-z's.

The input fluxes are also combined with their observational errors,
which are used for weighting each residual between the template and
observed fluxes in the EAZY fitting algorithm \citep{Brammer+08}. The supplementary information
on the uncertainty of each photometric measurement can enhance the
robustness of the colour-redshift mapping predicted by photo-z networks
\citep{Zhou+21}. The number of input variables $N_{\textrm{input}}$ is thus
twice the number of observational filters $N_{\textrm{filter}}$, with $N_{\textrm{input}}$
= 76, 70 and 52 for CDFS, COSMOS and UDS, respectively.

Our ML approach is to cast the photo-z estimation task into a classification
problem by binning the target redshift range into discretised classes
and returning a list of probabilities by which an example is found
in a given target bin. Multiple-bin regression has been used with template fitting methods in the past, but the benefit of this approach
has been demonstrated in recent ML photo-z studies \citep{Pasquet-Itam_Pasquet18,Pasquet+19,Lee_Shin21},
generally improving the photo-z accuracy. In the context of a model's
development, the probabilistic scrutiny of the redshift PDF allows
one to explore the causes of poor performance on some specific objects.
Reproducing realistic redshift PDFs as well as single-point estimates
could potentially contribute to improving cosmological analyses \citep[e.g.,][]{Mandelbaum+08,Myers+09,Palmese+20}.

Each PDF produced by our ML models is an output of the softmax function, which contains probabilities in $\zeta=\log(1+z)$ classes with a uniform distribution within $0<\zeta\lesssim1.8$, corresponding to the redshift range $0<z<5$.
The resolution of $\zeta$ bins approximates the PDF of $z$ provided by EAZY as the output vector.
The configuration adopted by \cite{Straatman+16} lets the algorithm explore a grid of redshifts with a step of $w_{z}=0.005(1+z)$.
The constant $\zeta$ bin width can be thus expressed as $w_{\zeta}\sim w_{z}/(1+z)=0.005$, which leads the photo-z network to output a vector of 350 probabilities as a PDF of $\zeta$ in our target redshift range.

\subsection{Optimisation of a baseline FCNN model}

\label{subsec:baseline_model}

We select a fully connected neural network (FCNN) as a baseline model, since it is commonly applied in photo-z estimation works.
Tables \hyperref[tab:major_fcnns]{B1} and \hyperref[tab:fcnns_w_simulations]{B2}
summarise some previous works which apply FCNNs to photo-z estimation, where the network was trained on spectroscopic samples in most cases.
This requires a huge amount of observational data and consequently results in a limited target redshift range up to no more than $\sim1-2$.
The number of filter bands used for photometric data is seldom as many as $\sim10$ as well, since cross-matching multiple catalogues tends to significantly reduce the sample size.

The updated S16 contains a much larger amount of photometric information with $\sim40$ filter bands, while our simulation method allows for training networks on sufficient mock data in a broader redshift range up to 5.
The architecture of the baseline FCNN should thus reflect the larger-scale configuration with more trainable parameters.
Other relevant works that have introduced photo-z based ML models trained with simulations typically adopt huge networks consisting of many layers and neurons: for example, $\{N_{\textrm{input}}:600:400:250\times13:N_{\textrm{output}}\}$ in \cite{Eriksen+20} and $\{N_{\textrm{input}}:512:1024:2048:1024:512:256:128:64:32:N_{\textrm{output}}\}$ in \cite{Ramachandra+22}.
We perform k-fold cross validation to explore the most appropriate architecture and optimise its hyperparameters by training the models on the simulated data generated with the Empirical noise model (iv), as described in $\S$ \ref{subsec:noise_model}.

Our photo-z code is designed for classifying input photometric data into 350 $\zeta$ bins, providing the output vector that represents a PDF of $\zeta$.
We thus employ the standard categorical cross-entropy (CCE) loss function \citep{Baum_Wilczek87,Solla+88}
\begin{equation}
L_{\textrm{CCE}}=-\sum_{c=1}^{C}y_{c}\log(s_{c}),\label{eq:L_CCE}
\end{equation}
where $y_{c}$ and $s_{c}$ are the ground truth and the score returned by the softmax function for each class $c$.
The redshift classifier is tuned so that the $\zeta$-prediction accuracy is maximised and the loss is minimised using one-hot encoding with $y_{c}=1$ only for a true class.

\begin{figure*}
\begin{centering}
\includegraphics[width=0.8\textwidth]{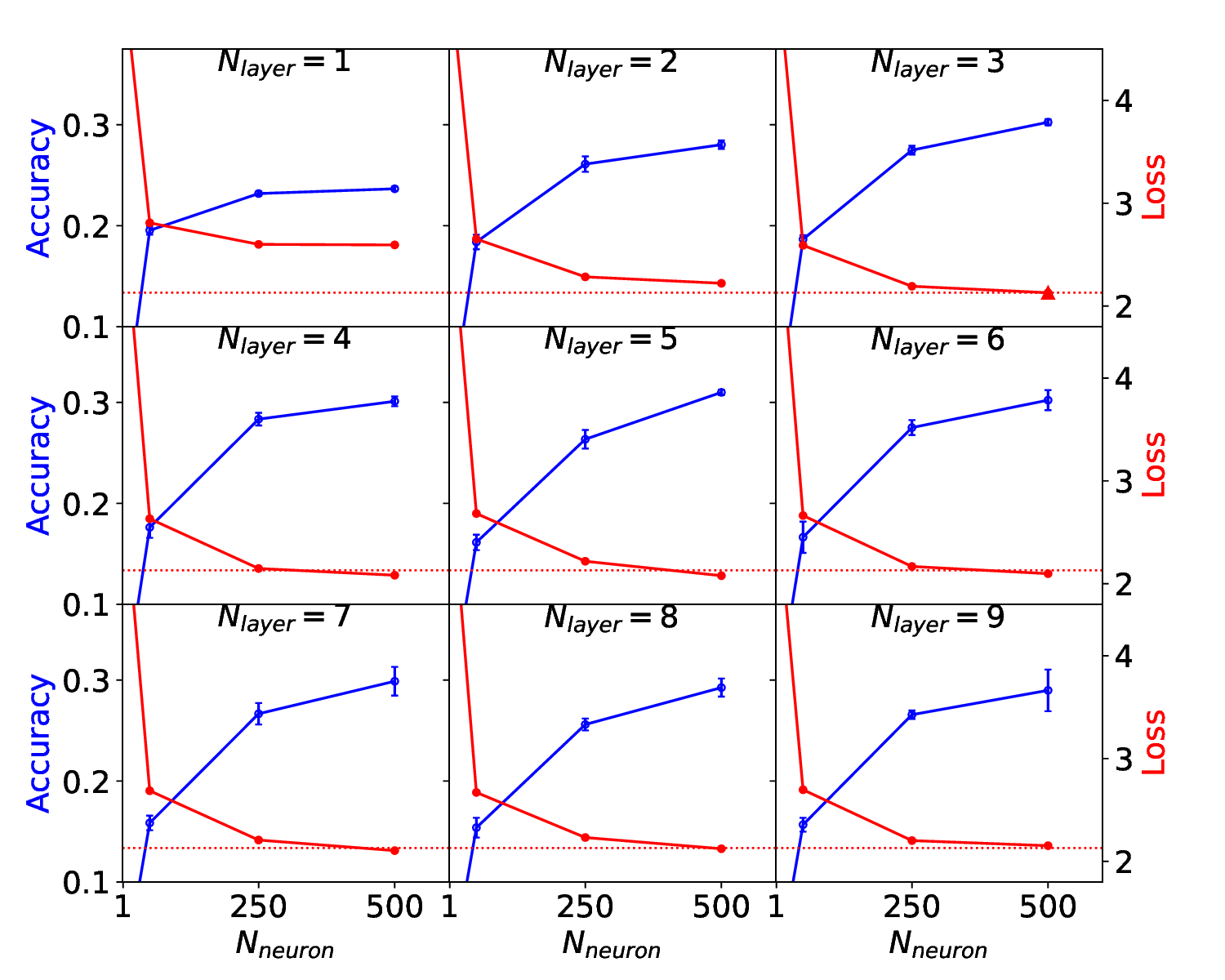}
\par\end{centering}
\caption{Optimisation of architecture and hyperparameters for the FCNN models
using 4-fold cross-validation. Each panel presents changes in validation
accuracy with the number of neurons ($N_{\textrm{neuron}}$) as the blue circles
for a given number of layers ($N_{\textrm{layer}}$). The accuracy score along
with its estimation error is given by the mean and standard deviation of the validation accuracy over all folds. The validation loss is also
shown by the red circles. The dotted horizontal line represents the
validation loss obtained from the baseline model, comprising 3 layers with 500 neurons each, which is presented
by the red triangle.\label{fig:cv_fcnn}}
\end{figure*}

For each FCNN we consider two types of hyperparameters relating to the architecture, the number of layers ($N_{\textrm{layer}}$) and the number of neurons in each layer ($N_{\textrm{neuron}}$), as well as those relating to the algorithm, namely learning rate and the dropout rate.
The latter (algorithmic) parameters are thus optimised for each set of the architectural ones. 
Fig. \ref{fig:cv_fcnn} shows the results on hyperparameter optimisation for the FCNN, presenting the validation accuracy and loss for each combination of $N_{\textrm{layer}}$ and $N_{\textrm{neuron}}$ within the ranges $N_{\textrm{layer}}\in[1,9]$ and $N_{\textrm{neuron}}\in[1,500]$.
The accuracy is defined as the percentage of predicted redshift classes that match with true ones. Note, we don't expect accuracy
to reach 100\% even when performing well, since we expect scatter into neighbouring redshift bins as photo-zs are intrinsincally uncertain, and some redshifts will lie closer to the bin boundaries. Nevertheless, for a fixed validation sample, it is a good relative indicator. We explore
other metrics below.

Each panel presents changes in accuracy scores with $N_{\textrm{neuron}}$ for a given $N_{\textrm{layer}}$.
We find that the accuracy levels off with increasing $N_{\textrm{neuron}}$ if the individual layers contain sufficient neurons.
This is not affected by the number of layers in general with the accuracy converging to $\gtrsim30\%$.
The minimum loss can be attained by the model with $(N_{\textrm{neuron}},N_{\textrm{layer}})=(500,3)$, with no significant improvement from increasing the number of trainable parameters with larger $N_{\textrm{neuron}}$ or $N_{\textrm{layer}}$.
The architecture of our FCNN model is therefore constructed from three layers with 500 neurons, since a smaller architecture is preferable to a larger one for the same performance.
The number of weights to be trained is $\sim700,000$.

\begin{figure*}
\begin{centering}
\includegraphics[width=1\textwidth]{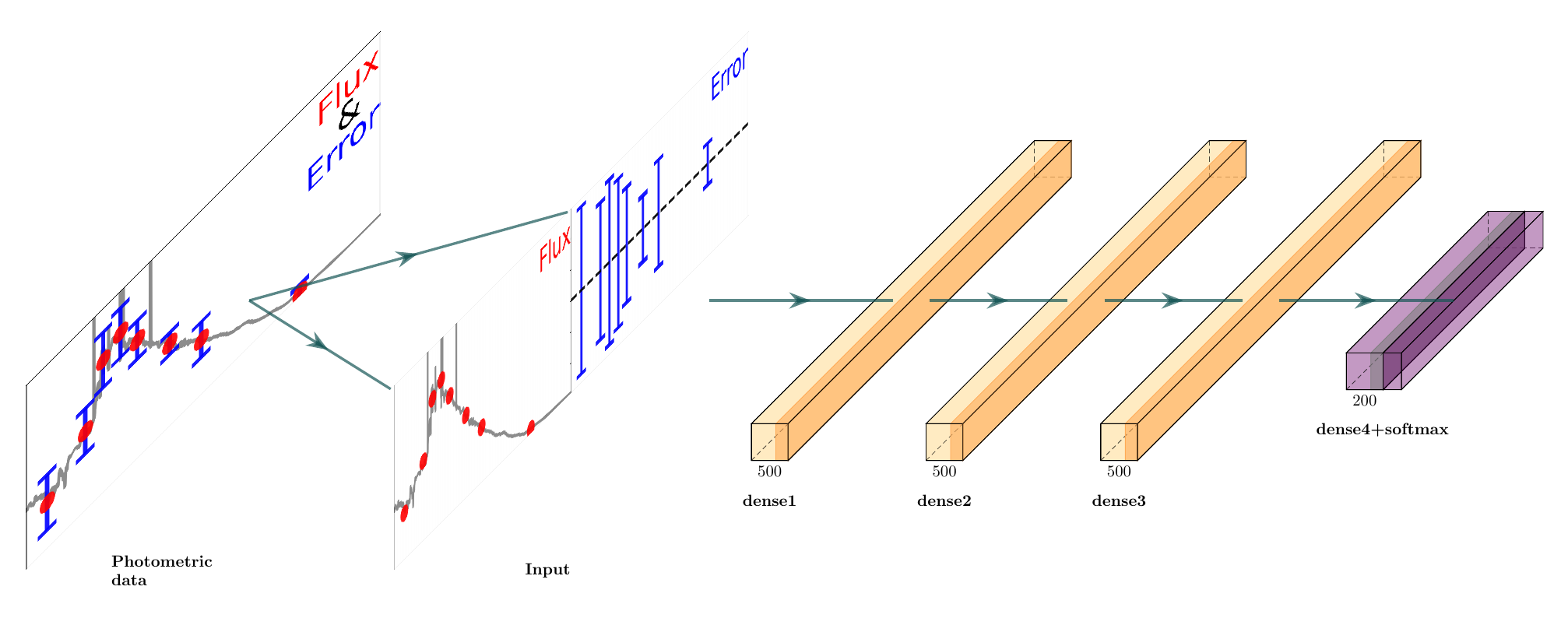}
\par\end{centering}
\caption{The network architecture of the baseline FCNN classifier. Each figure
indicates the output dimension and the model consists of 3 layers
with 500 neurons (yellow). Following the intermediate linear layers
are ReLU non-linearities, 5 per cent dropout and batch normalisation
layer (orange). Galaxy flux ratios coupled with normalised observational
errors are fed into the photo-z network, which provides the softmax
output (purple) of 350 probability scores for discretised $\zeta=\log(1+z)$
bins. \label{fig:architecture_of_fcnn}}
\end{figure*}

Fig \ref{fig:architecture_of_fcnn} visualises the overall architecture
of the optimised baseline model with some details excluded. Each layer
is followed by ReLU non-linearities, 5 per cent dropout and a batch
normalisation layer. The input flux ratios along with their observational
errors are fed into the network with missing values included, which
produces the softmax output of 350 $\zeta$ probabilities.

In the initial exploratory phase of this research other ML techniques were also tested, using a similar hyperparameter optimisation strategy.
The performance of random forests (RFs) and support vector machines (SVMs) was examined with different sets of hyperparameters: the number of estimators and max depth for RFs and ($C,\gamma$) for SVMs, where $C$ controls the complexity of the decision surface while $\gamma$ the range influenced by a single data point.
Each model was developed with its best hyperparameters, but underperformed the FCNN in that their validation accuracies only reached just under 30\%.
This indicates that neural networks are more appropriate for our photo-z estimation scheme than other major ML approaches. In particular, with neural networks we have the ability to optimise the loss function for PDF recovery (see discussed in \ref{subsec:PDF-statistics}).

\subsection{Architecture of HAYATE}

\label{subsec:Architecture-of-HAYATE}

We further develop a CNN-based photo-z network and compare the performance
of these different ML approaches. As before, the output is a probability vector
on discretised redshift bins, which translates the regression problem
into a classification task and provides redshift PDFs as well as their
point estimates. The output PDF is produced by combining multiple
networks independently trained with different configurations, representing
an ensemble of stochastic variants for each test object.

We build HAYATE with the CNN architecture inspired by the VGG neural
network \citep[VGGNet;][]{Simonyan_Zisserman15}, one of the simplest
CNN structures commonly used for image classification and object detection.
The extended variant of the VGG model, VGG19, consists of 16 convolution
layers with 5 max pooling layers followed by 3 fully connected layers
and 1 softmax layer. It features an extremely small receptive field,
a kernel of $3\times3$, which is the smallest size that can capture
the neighbouring inputs. Stacking multiple $3\times3$ convolutions
instead of using a larger receptive field leads to a deeper network,
which is required for better performance \citep{EmmertStreib+20}.
VGG-based models have been successfully applied to astronomical images,
e.g. for the identification of radio galaxies \citep{Wu+19},
classification of compact star clusters \citep{Wei+20} and detection
of major mergers \citep{Wang+20}.

The VGG network is fundamentally designed for handling higher-dimensional image data (with multiple colour channels) rather than 1D photometric data.
It should be thus applied to photo-z estimation with a much smaller architecture, since the number of trainable parameters originally reaches up to $\sim144$ million.
\cite{Zhou+21} have introduced a 1D CNN used for deriving spec-z's from spectral data, which can provide some insight into the application of CNNs to photo-z estimation.
The input layer includes two channels of spectral data and errors, while the output layer contains multiple neurons representing the probability of each redshift interval.
The spec-z analysis is thus performed as a classification task using the feature maps obtained through two convolutional layers, which are followed by two fully connected layers.
The number of parameters is consequently far less than that of a CNN commonly used for image processing, totalling no more than $\sim350,000$.

\begin{figure*}
\begin{centering}
\includegraphics[width=1\textwidth]{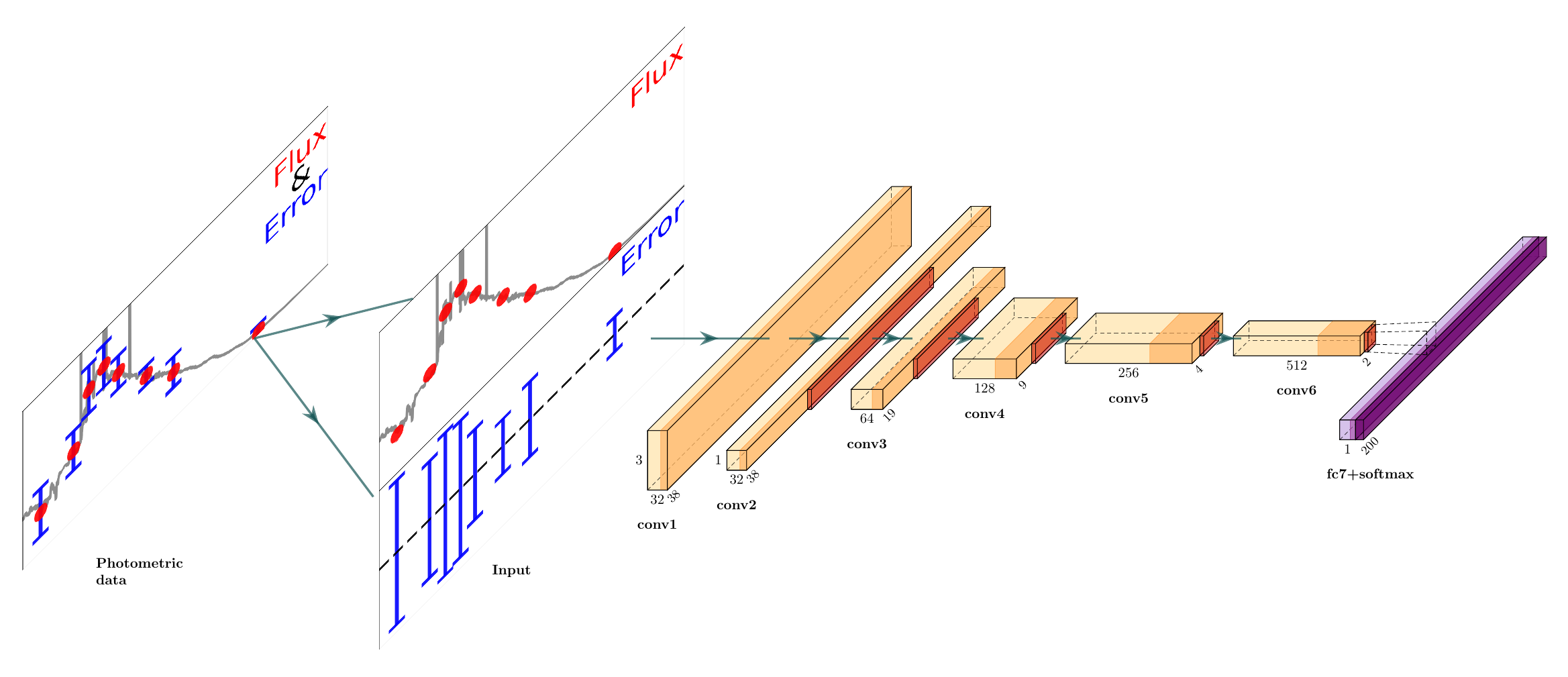}
\par\end{centering}
\caption{Architecture of HAYATE. The overall structure consists of 6 convolutional
layers (yellow) with 32, 32, 64, 128, 256 and 512 kernels each. The input
$2\times N_{\textrm{filter}}$ matrix contains 2 rows of flux ratios and normalised
observational errors. In the first layer, we convolve the input with
a kernel of $2\times3$ using zero padding of size 1. The convolution
operation in the following layer then performs on the $3\times N_{\textrm{filter}}$
matrix with a kernel of $3\times3$, outputting a 1D vector of size
$N_{\textrm{filter}}$ by adopting zero padding in the column direction. Each
remaining layer is set with a 1D kernel of size 3, which reflects
the fundamental concept of the VGG network. All the convolutional
layers are connected with batch normalisation, dropout (orange) and
1D max pooling (dark orange) layers except for the first one that
lacks a max pooling operation. We set the last layer to a fully connected
output layer with $350$ neurons (purple), each providing a softmax
probability score for a given $\zeta$ class. \label{fig:architecture_of_hayate}}
\end{figure*}

The task of photo-z prediction can be treated in the same fashion but with input flux and output probability vectors of lower resolution than spec-z.
We construct HAYATE as a simplified variant of the VGG network, whose architecture is illustrated in Fig. \ref{fig:architecture_of_hayate}.
The input $2\times N_{\textrm{filter}}$ matrix involves 2 rows of flux ratios and normalised observational errors, convolved with a kernel of $2\times3$ using zero padding of size 1 and  followed by the $3\times N_{\textrm{filter}}$ matrix.
Adopting zero padding in the column direction, we then convolve it with a kernel of $3\times3$ to obtain a 1D vector of size $N_{\textrm{filter}}$.
The major components are a following sequence of 6 convolutional layers with 32, 32, 64, 128, 256 and 512 kernels each.
The fundamental concept of the VGG network is particularly reflected by a 1D kernel of size 3 used for a convolution operation in each layer.
We basically connect the convolutional layers with batch normalisation, dropout and 1D max pooling layers.
A fully connected layer is set with $350$ neurons in the end, each outputting a softmax probability of finding an object at a given $\zeta$.

We have explored more efficient architectures in supplementary experiments, only to conclude that the one mentioned above should perform best among several simple CNNs.
We also note that the number of trainable weights for HAYATE is approximately the same as that of the baseline FCNN described in \S \ref{subsec:baseline_model}.

\subsection{Evaluation metrics}

\label{subsec:Indicators-for-photo-z-estimates}

All the ML photo-z's are estimated from the redshift PDFs in the same
method as implemented in EAZY by \cite{Straatman+16}. Each point
estimate is obtained by marginalizing exclusively over the peak of
the redshift PDF which shows the largest integrated probability. This
adapts to the degeneracy of template colours with redshift, which
produces a PDF with multiple peaks.

The quality of photo-z estimates is evaluated based on the residuals
with respect to their spec-z's, which are given by
\begin{eqnarray}
\Delta z & = & \frac{z_{\textrm{phot}}-z_{\textrm{spec}}}{1+z_{\textrm{spec}}},\label{eq:delta_z}
\end{eqnarray}
where $z_{\textrm{phot}}$ and $z_{\textrm{spec}}$ are photometric and spectroscopic
redshifts. Each ML photo-z is immediately recovered from a point estimate of $\zeta$, expressed as $z_{\textrm{phot}}=e^{\zeta}-1$. We employ the following commonly used indicators as statistical
metrics to evaluate the model's performance in single-point estimations:
\begin{itemize}
\item $\sigma_{\textrm{NMAD}}$: normalised absolute median deviation of $\Delta z$,
described as
\begin{eqnarray}
\sigma_{\textrm{NMAD}} & = & 1.48\times median(|\Delta z|),\label{eq:sigma_nmad}
\end{eqnarray}
which is robust to $\Delta z$ outliers.
\item Outlier rate $\eta_{0.2}$: percentage of outliers, defined as test
data with $|\Delta z|>0.2$.
\end{itemize}
We also use the probability integral transform \citep[PIT;][]{Polsterer+16}
to properly estimate the calibration of the redshift PDF $P(\zeta)$ generated
by different photo-z models, which is defined by
\begin{eqnarray}
x_{t}= PIT_{t}(\zeta_{\textrm{spec},t}) & = & \int_{-\infty}^{\zeta_{\textrm{spec},t}}P_{t}(\zeta)d\zeta,\label{eq:pit}
\end{eqnarray}
where $\zeta_{\textrm{spec},t}$ corresponds to the true redshift of the test source $t$. If the
predicted PDFs are well calibrated with respect to the spec-z's, the
histogram of the PIT values, or its PDF $f(x)$, is equivalent to
the uniform distribution $U(0,1)$. The flat distribution indicates
that the predicted PDFs are neither biased, too narrow nor too broad.
Conversely, underdispersed and overdispersed PDFs exhibit U-shaped and centre-peaked
distributions, respectively, while a systematic bias present in the
PDFs is represented by a slope in the PIT distribution.

\begin{table*}
\begin{centering}
\caption{Summary of evaluation metrics. We employ $\sigma_{\textrm{NMAD}}$ and $\eta_{0.2}$
for measuring the accuracy of photo-z point estimates, while KL, CvM
and CRPS are responsible for assessing the quality of photo-z PDFs.
\label{tab:indicators}}
\par\end{centering}
\centering{}%
\begin{tabular}{ccccc}
\hline 
\multirow{1}{*}{Indicator} & Target to be evaluated & Responsibility & Key attribute & Measurement\tabularnewline
\hline
\hline
$\sigma_{\textrm{NMAD}}$ & \multirow{2}{*}{$z_{\textrm{phot}}$ point estimate}  & $z_{\textrm{phot}}$ precision & \multirow{2}{*}{$\Delta z$ {[}Eq. (\ref{eq:delta_z}){]}} & Median of $|\Delta z|$\tabularnewline
$\eta_{0.2}$ &  & Rate of catastrophically wrong $z_{\textrm{phot}}$ &  & Outlier rate of $\Delta z$ with $|\Delta z|>0.2$\tabularnewline
\hline
\multirow{1}{*}{KL} & \multirow{3}{*}{$P(z_{\textrm{phot}})$} & \multirow{2}{*}{Calibration of produced $P(z_{\textrm{phot}})$} & \multirow{2}{*}{$x$: PIT {[}Eq. (\ref{eq:pit}){]}} & Divergence of PIT distribution $f(x)$ from uniformity\tabularnewline
\multirow{1}{*}{CvM} &  &  &  & Dissimilarity between CDF of $f(x)$ and identity line\tabularnewline
CRPS &  & Reliability of $P(z_{\textrm{phot}})$ w.r.t. $z_{\textrm{spec}}$ & $crps$ {[}Eq. (\ref{eq:crps_i}){]} & Median of $crps$\tabularnewline
\hline 
\end{tabular}
\end{table*}

The following evaluation metrics are used for quantifying the global
property of output PDFs:
\begin{itemize}
\item CvM: score of a Cram\'{e}r-von Mises \citep{Cramer28} test,
\begin{equation}
CvM=\int_{-\infty}^{\infty}[F(x)-F_{U}(x)]^{2}dF_{U},\label{eq:cvm}
\end{equation}
where $F(x)$ and $F_{U}(x)$ are cumulative distribution functions (CDFs) of $f(x)$ and $U(0,1)$, respectively.
This corresponds to the mean-squared difference between the CDFs of
the empirical and true PDFs of PIT.
\item KL: Kullback--Leibler \citep{Kullback_Leibler51} divergence,
\begin{equation}
KL=\int_{-\infty}^{\infty}f(x)ln\left(\frac{U(0,1)}{f(x)}\right)dx,\label{eq:kl}
\end{equation}
which is a statistical distance representing the information loss
in using $f(x)$ to approximate $U(0,1)$. An approximation $f(x)$
closer to $U(0,1)$ thus shows a smaller KL.
\end{itemize}
The reliability of individual PDFs with respect to spec-z's is represented
by the Continuous Ranked Probability Score \citep[CRPS;][]{Hersbach00,Polsterer+16},
which is given by
\begin{equation}
\mathrm{\textit{crps}}_{t}=\int_{-\infty}^{\infty}[C_{t}(\zeta)-C_{\textrm{spec},t}(\zeta)]^{2}d\zeta,\label{eq:crps_i}
\end{equation}
where $C_{t}(\zeta)$ and $C_{\textrm{spec},t}(\zeta)$ are CDFs of $P_{t}(\zeta)$ and $\zeta_{\textrm{spec},t}$ for the source $t$,
respectively. $C_{\textrm{spec},t}(\zeta)$ here corresponds to the CDF of $\delta(\zeta-\zeta_{\textrm{spec},t})$
\begin{equation}
C_{\textrm{spec},t}(\zeta)= H(\zeta-\zeta_{\textrm{spec},t}),\label{eq:heaviside}
\end{equation}
where $H(\zeta-\zeta_{\textrm{spec},t})$ is the Heaviside step-function, which gives
0 for $\zeta<\zeta_{\textrm{spec},t}$ and 1 for $\zeta\geq \zeta_{\textrm{spec},t}$. It reflects the simplest form for the unknown true distribution of $\zeta$, given by the Dirac Delta function between $\zeta$ and $\zeta_{\textrm{spec},t}$. The CRPS thus
represents the distance between $C_{t}(\zeta)$ and $C_{\textrm{spec},t}(\zeta)$,
or the difference between the empirical and ideal redshift PDFs. We
thus assess the reliability of individual output PDFs with the median
value of $crps$, which is robust to outliers than their mean value:
\begin{itemize}
\item CRPS: median of all CRPS values obtained from a sample, 
\begin{equation}
\mathrm{CRPS}=\underset{t}{\mathrm{median}}(\textit{crps}{}_{t}).\label{eq:crps}
\end{equation}
\end{itemize}
We introduced the CRPS metric primarily because it can be used as part of the ANN optimisation process.

Table \ref{tab:indicators} summarises the characteristics of these
indicators. In summary, $\sigma_{\textrm{NMAD}}$ and CRPS reveal the quality of
individual outputs while $\eta_{0.2}$, KL and CvM represent the global
property obtained from the distribution of their key attributes.

\subsection{Training process}

\label{subsec:Training-process}

The mock photometric data are divided into three parts representing training, validation and test datasets.
The test sample contains 20\% of the whole set of simulated galaxies, while the rest is split into the training and validation sets with 70\% and 30\% of the remaining data randomly selected, respectively.
The individual networks are trained with a joint loss that combines the CCE loss $L_{\textrm{CCE}}$ and the CRPS loss $L_{\textrm{CRPS}}$, given by equations (\ref{eq:L_CCE}) and (\ref{eq:crps_i}), respectively.
The CCE loss is frequently used for multi-class classification problems, responsible for the accuracy of single point estimates.
The CRPS loss can function as a penalty for failing to produce reliable PDFs, which would be otherwise neglected in a classification task with the single CCE loss.

The joint loss is optimised by an Adam optimiser \citep{Kingma_Ba15} in the training process, which is given by
\begin{equation}
L=\alpha L_{\textrm{CCE}}+\beta L_{\textrm{CRPS}},\label{eq:compound_loss}
\end{equation}
where $\alpha$ and $\beta$ are the weights to the linear combination of the CCE and CRPS losses.
We first explore the appropriate values for $\alpha$ and $\beta$ in a pre-training process so that $L_{\textrm{CCE}}$ and $L_{\textrm{CRPS}}$ equivalently contribute to the total at loss convergence with $\alpha L_{\textrm{CCE}}\simeq\beta L_{\textrm{CRPS}}$.
This is achieved by updating them after each training epoch $j$ with the following equations:
\begin{align}
\alpha_{j+1} & =\frac{L_{j}}{2L_{\textrm{CCE},j}},\label{eq:alpha}\\
\beta_{j+1} & =\frac{L_{j}}{2L_{\textrm{CRPS},j}},\label{eq:beta}
\end{align}
where $\alpha_{j+1}$ and $\beta_{j+1}$ are the updated coefficients used for the next training step.
The training terminates with an early stopping method after 10 epochs of no improvement in model's performance on the held-out validation set.
We then train the network from scratch using the fixed coefficients of the convergence for $\alpha$ and $\beta$ which are obtained from the pre-training.

\subsection{Transfer learning}

\label{subsec:Transfer-learning}

We apply transfer learning to HAYATE to build an empirically fine-tuned
model (HAYATE-TL) and see if it can exploit the spec-z information.
In transfer learning, typically the last layers of a pre-trained network are re-trained on a different dataset while the rest of the layers' weights remain frozen (fixed). A
pre-trained model is a saved network that has been trained
on a large dataset, then learns new features from a distinct training sample
in another domain or regime.
Here we fine-tune the last two convolutional layers which have been trained on the simulated
datasets with the observed samples with spec-z information. It should
be noted that re-training more layers does not show a significant improvement (in this case, and in general - see \citet{transfer})
in the model's performance and we thus allow only the last two layers
to be trainable with spectroscopic observations.

The estimated photo-z's of a test sample are to some degree dependent on which particular SEDs are included in the training and test sets.
We implement a method of building
a robust estimator by combining multiple PDFs for each object, which
are produced by distinct models whose training sets do not contain
the same object. The entire spec-z catalogue is sliced into 90\% and
10\% for training and test sets, respectively, which is repeatedly
performed to provide 50 different 90-10 splits. Each test set source
is then included in 5 different test samples, whose corresponding
training sets are used for optimising 15 lower-level networks in the
framework of ensemble learning, as discussed in \S \ref{subsec:Ensemble-learning}.
The output PDF obtained with transfer learning thus results in a combination
of 75 different PDFs provided for each object.

\subsection{Ensemble learning}

\label{subsec:Ensemble-learning}

Randomness appears in many aspects of the training process, which
makes the weights of the model converge to different local minima
of the loss function even if the datasets used are exactly the same.
Prior to the training, splitting a dataset into training, validation
and test sets is often done randomly depending on each experiment.
The initial values of the weights are also randomised so that the
training processes start with different initial states. During the
training, the shuffled batches also lead to different gradient values
across runs, while a subset of the neurons are randomly ignored by the
dropout layers.

The effect of local minima can be reduced by performing Bootstrap
AGGregatING \citep[Bagging,][]{Breiman96,Dietterich97}, which integrates
multiple models trained on different datasets that are constructed
by sampling the same training set with replacement. The main principle
behind the bagging algorithm is to build a generic model by combining
a collection of weak learners that are independently trained with
the uncorrelated subsets from the original training set. The composite
strong learner can outperform a single model established on the original
sample \citep{Rokach10}.

A random forest ensemble \citep{Breiman_Schapire01} is commonly adopted
in the field of ensemble learning, which is characterised by a number
of decision trees, each trained on a different subset of the entire
training sample. The benefit obtained from these techniques has been
demonstrated for a wide range of regression and classification tasks
in astronomy \citep[e.g.,][]{Way+06,CarrascoKind_Brunner14,Kim+15,Baron_Poznanski17,Green+19}.
Some ML photo-z studies have succeeded in applying the construction
of prediction trees and the RF techniques to improve the redshift
estimation accuracy. \citep{CarrascoKind_Brunner13,Cavuoti+17}.

\begin{figure*}
\begin{centering}
\includegraphics[width=0.8\textwidth]{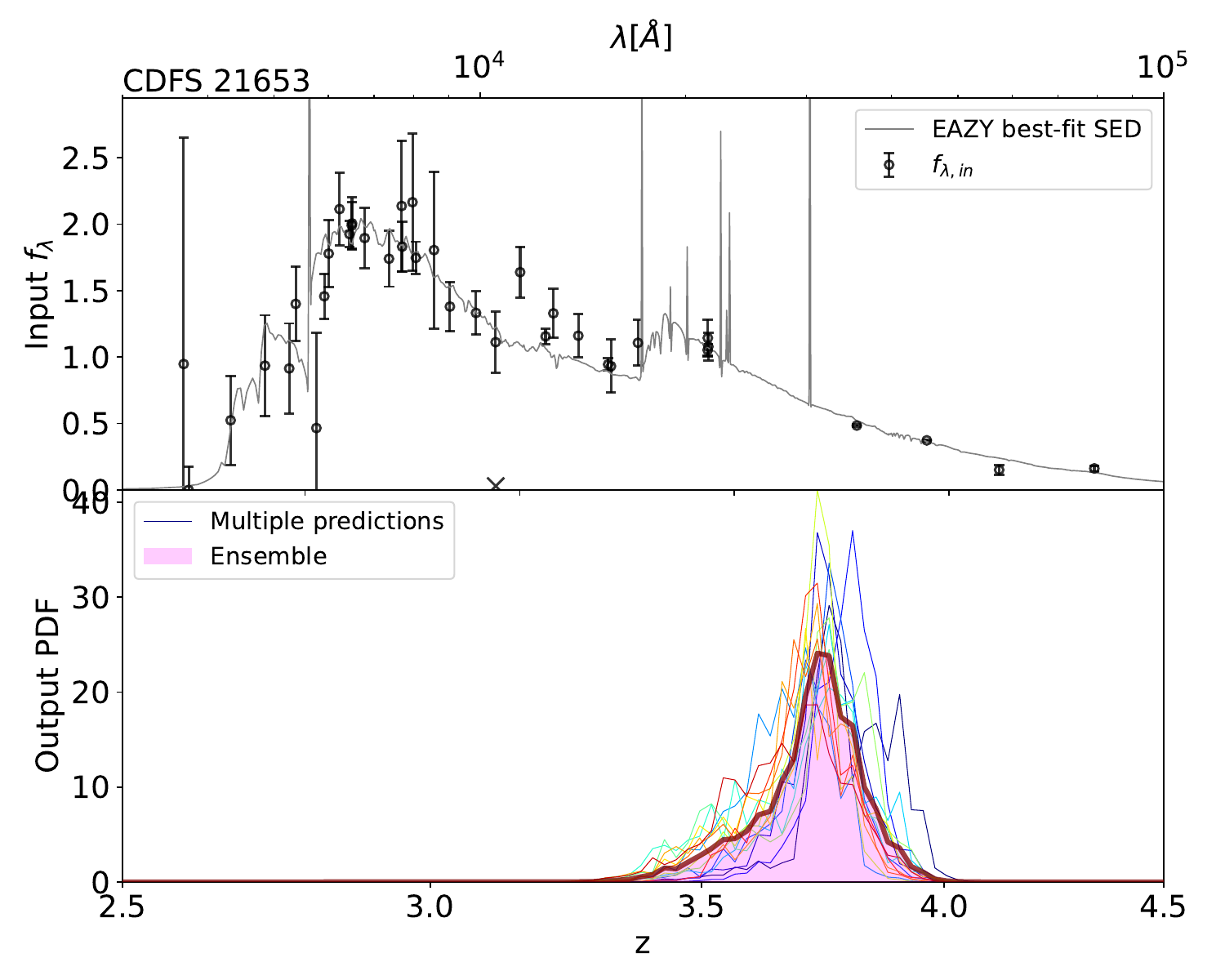}
\par\end{centering}
\caption{Top: inputs of fluxes and observational errors for an example object
in CDFS. The normalised fluxes and photometric errors are presented
by the black circles with error bars. The gray line shows the corresponding
best-fit SED derived with EAZY. Bottom: ensemble of output PDFs as
a function of $\zeta=\log(1+z)$, shown by the shaded region coloured
in purple. The solid lines in different colours are lower-level PDFs
produced by 15 different networks for the source presented in the
top panel, which are combined into the thick purple line as an ensemble.
\label{fig:combine_pdfz}}
\end{figure*}

We rather use a smaller subset of the full simulated data for training
each network, instead of generating a bootstrapped sample of full
sample size $N_{\textrm{train}}$. The training sub-samples are constructed by partitioning
the full data into 3, which ensures the independence of each subset
while the training is computationally less intensive due to the smaller
sample size. We thus train each network on a sub-sample of size $N_{\textrm{train}}/3$,
obtained from the 5 individual training sets of different noise realisations.
The ensemble of multiple PDFs $P_{i,j}(\zeta)$ is thus given by
\begin{equation}
P(\zeta)=\frac{1}{15}\sum_{i}^{5}\sum_{j}^{3}P_{i,j}(\zeta),\label{eq:ensemble_P}
\end{equation}
where $i$ is the index of the simulated dataset of different noise
realisation while $j$ discriminates the sub-samples. It follows the
output PDF of each sample galaxy is produced by averaging 15 lower-level
predictions, whose typical example is shown in Fig. \ref{fig:combine_pdfz}.
This allows for outputting more robust and reliable PDFs than those
obtained with a single network \citep{Sadeh+16,Eriksen+20}.

\section{Results}

\label{sec:results}

\renewcommand{\arraystretch}{1.5}

\begin{table*}
\begin{centering}
\caption{Performance of different photo-z models on the spec-z samples provided by S16.
The number of inputs ($N_{\textrm{input}}$) and the sample size of spec-z data ($N_{\textrm{spec}}$) are presented in the second and third columns, for CDFS, COSMOS and UDS from top to bottom.
For each field, the photo-z and PDF statistics are shown for the baseline FCNN, HAYATE, HAYATE-TL and EAZY.
We additionally train the CNN of the same architecture as HAYATE purely with the spec-z data from scratch to exhibit the benefit of training with simulations.
All the uncertainties are the standard deviation derived from bootstrap resampling. \label{tab:results_tab}}
\par\end{centering}
\centering{}%
\scalebox{0.9}[0.9]{
\begin{tabular}{cccccccccc}
\toprule 
\multirow{2}{*}{Field} & \multirow{2}{*}{$N_{\textrm{input}}$} & \multirow{2}{*}{$N_{\textrm{spec}}$} & \multirow{2}{*}{Photo-z model} & \multirow{2}{*}{Photo-z method} & \multicolumn{5}{c}{Evaluation metric}\tabularnewline
 &  &  &  &  & $\sigma_{\textrm{NMAD}}[10^{-2}]$ & $\eta_{0.2}[\%]$ & $\mathrm{KL}[10^{-1}]$ & $\mathrm{CvM}[10^{-3}]$ & $\mathrm{CRPS}[10^{-4}]$\tabularnewline
\midrule
\multirow{5}{*}{CDFS} & \multirow{5}{*}{$38\times2$} & \multirow{5}{*}{1274} & \textbf{EAZY} & \textbf{Template fitting} & \textbf{$\mathbf{1.14_{-0.06}^{+0.05}}$} & \textbf{$\mathbf{1.26_{-0.31}^{+0.31}}$} & \textbf{$\mathbf{0.96\pm0.12}$} & \textbf{$\mathbf{3.60_{-0.59}^{+0.92}}$} & \textbf{$\mathbf{0.88_{-0.05}^{+0.06}}$}\tabularnewline
 &  &  & \textbf{HAYATE} & \textbf{Trained with simulations} & \textbf{$\mathbf{0.96_{-0.05}^{+0.05}}$} & \textbf{$\mathbf{0.94_{-0.24}^{+0.24}}$} & \textbf{$\mathbf{0.45\pm0.09}$} & \textbf{$\mathbf{2.32_{-0.70}^{+0.96}}$} & \textbf{$\mathbf{0.83_{-0.03}^{+0.04}}$}\tabularnewline
 &  &  & HAYATE-TL & Transfer learning with observations & $0.74_{-0.02}^{+0.03}$ & $0.94_{-0.24}^{+0.24}$ & $0.35\pm0.09$ & $1.90_{-0.56}^{+0.77}$ & $0.70_{-0.04}^{+0.06}$\tabularnewline
 &  &  & FCNN & Trained with simulations & $1.26_{-0.06}^{+0.05}$ & $1.41_{-0.31}^{+0.31}$ & $0.38\pm0.08$ & $1.14_{-0.40}^{+0.63}$ & $1.55_{-0.08}^{+0.08}$\tabularnewline
 &  &  & CNN & Trained purely with observations & $5.11_{-0.24}^{+0.29}$ & $2.75_{-0.47}^{+0.47}$ & $0.68\pm0.13$ & $1.84_{-0.38}^{+0.54}$ & $25.62_{-0.54}^{+0.82}$\tabularnewline
\midrule
\multirow{5}{*}{COSMOS} & \multirow{5}{*}{$35\times2$} & \multirow{5}{*}{738} & \textbf{EAZY} & \textbf{Template fitting} & \textbf{$\mathbf{1.53_{-0.11}^{+0.07}}$} & \textbf{$\mathbf{1.90_{-0.54}^{+0.54}}$} & \textbf{$\mathbf{0.78\pm0.15}$} & \textbf{$\mathbf{2.14_{-0.40}^{+0.89}}$} & \textbf{$\mathbf{1.72_{-0.10}^{+0.15}}$}\tabularnewline
 &  &  & \textbf{HAYATE} & \textbf{Trained with simulations} & \textbf{$\mathbf{1.42_{-0.06}^{+0.08}}$} & \textbf{$\mathbf{1.22_{-0.41}^{+0.41}}$} & \textbf{$\mathbf{0.42\pm0.12}$} & \textbf{$\mathbf{1.27_{-0.42}^{+0.89}}$} & \textbf{$\mathbf{1.76_{-0.15}^{+0.12}}$}\tabularnewline
 &  &  & HAYATE-TL & Transfer learning with observations & $1.26_{-0.11}^{+0.06}$ & $1.22_{-0.41}^{+0.41}$ & $0.34\pm0.11$ & $0.33_{-0.03}^{+0.49}$ & $1.48_{-0.12}^{+0.11}$\tabularnewline
 &  &  & FCNN & Trained with simulations & $1.54_{-0.05}^{+0.14}$ & $1.90_{-0.54}^{+0.54}$ & $0.57\pm0.13$ & $0.90_{-0.28}^{+0.73}$ & $2.32_{-0.11}^{+0.12}$\tabularnewline
 &  &  & CNN & Trained purely with observations & $7.72_{-0.35}^{+0.30}$ & $6.78_{-0.81}^{+0.95}$ & $1.76\pm0.29$ & $5.09_{-0.68}^{+1.08}$ & $59.28_{-2.10}^{+2.60}$\tabularnewline
\midrule
\multirow{5}{*}{UDS} & \multirow{5}{*}{$26\times2$} & \multirow{5}{*}{312} & \textbf{EAZY} & \textbf{Template fitting} & \textbf{$\mathbf{1.90_{-0.13}^{+0.09}}$} & \textbf{$\mathbf{1.28_{-0.64}^{+0.64}}$} & \textbf{$\mathbf{0.70\pm0.23}$} & \textbf{$\mathbf{2.37_{-0.51}^{+1.56}}$} & \textbf{$\mathbf{2.59_{-0.11}^{+0.15}}$}\tabularnewline
 &  &  & \textbf{HAYATE} & \textbf{Trained with simulations} & \textbf{$\mathbf{1.94_{-0.08}^{+0.07}}$} & \textbf{$\mathbf{0.32_{-0.32}^{+0.32}}$} & \textbf{$\mathbf{0.70\pm0.22}$} & \textbf{$\mathbf{2.84_{-0.85}^{+1.78}}$} & \textbf{$\mathbf{2.63_{-0.37}^{+0.28}}$}\tabularnewline
 &  &  & HAYATE-TL & Transfer learning with observations & $1.82_{-0.13}^{+0.08}$ & $0.32_{-0.32}^{+0.32}$ & $0.42\pm0.18$ & $1.52_{-0.54}^{+1.37}$ & $2.69_{-0.42}^{+0.31}$\tabularnewline
 &  &  & FCNN & Trained with simulations & $2.21_{-0.12}^{+0.09}$ & $0.64_{-0.32}^{+0.32}$ & $0.40\pm0.18$ & $1.78_{-0.71}^{+1.62}$ & $3.24_{-0.21}^{+0.24}$\tabularnewline
 &  &  & CNN & Trained purely with observations & $11.12_{-1.04}^{+0.95}$ & $15.38_{-2.24}^{+1.92}$ & $2.60\pm0.68$ & $6.65_{-0.85}^{+1.63}$ & $214.15_{-10.40}^{+12.72}$\tabularnewline
\bottomrule
\end{tabular}
}
\end{table*}

\renewcommand{\arraystretch}{1.}

We evaluate the performance of HAYATE on the spec-z samples in S16, particularly for CDFS and COSMOS, each containing $\sim1000$ galaxies with photometric data of $\lesssim40$ filters.
It is also tested on the smaller sample from UDS for a supplementary experiment, although no more than 312 objects are available with 26 input fluxes provided for each.
Table \ref{tab:results_tab} gives an overview of the results for EAZY, HAYATE and HAYATE-TL along with the baseline FCNN.
We also probe the benefit of learning from simulated data by training the CNN of the same architecture as HAYATE purely with the spec-z data from scratch.

\begin{figure*}
\begin{centering}
\includegraphics[width=0.9\textwidth]{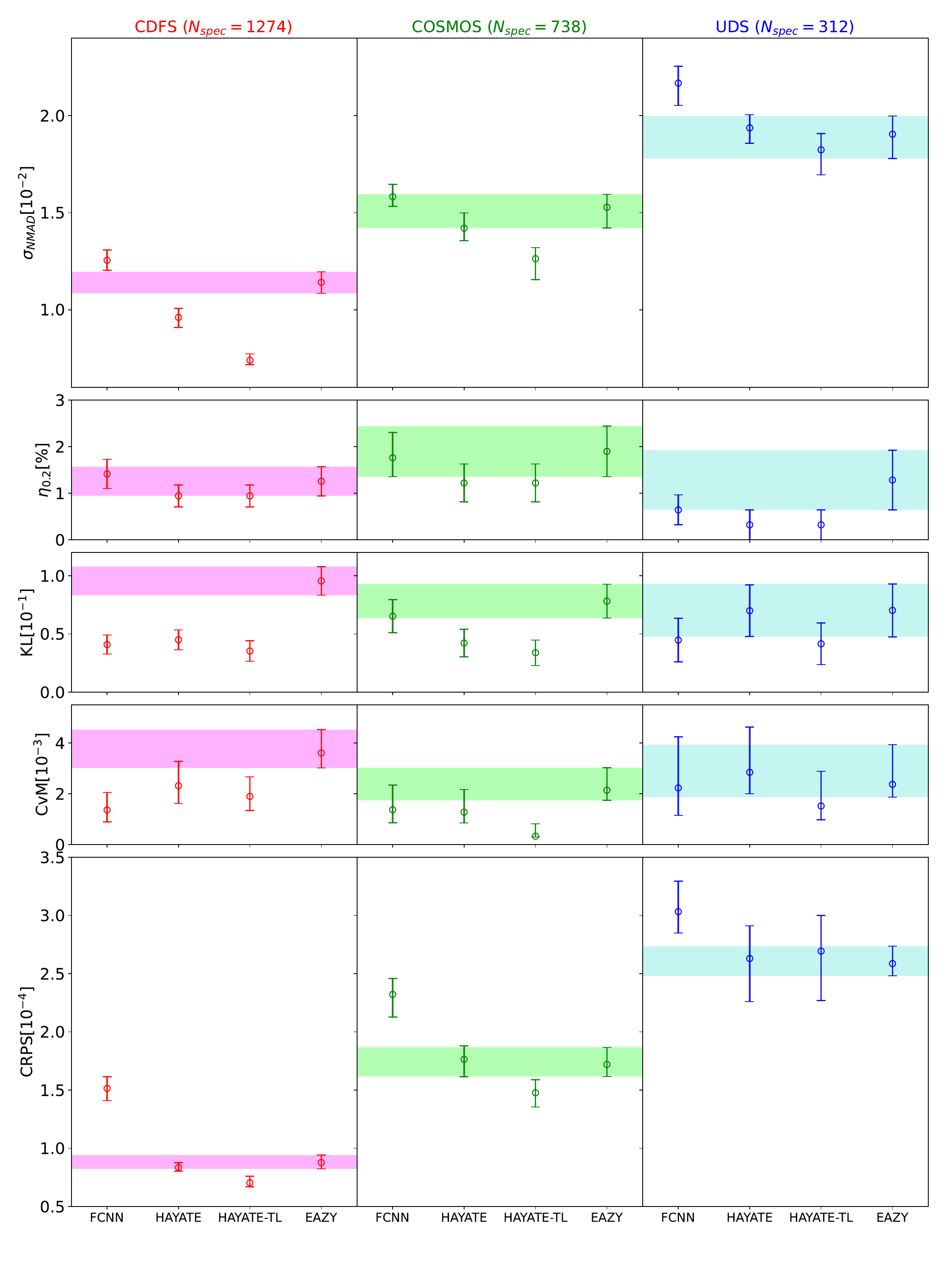}
\par\end{centering}
\caption{Visualisation of the comparison in the photo-z and PDF statistics
between different models presented in Table \ref{tab:results_tab}.
The results for CDFS, COSMOS and UDS are provided in the left, middle
and right columns, respectively, which are individually coloured in
red, green and blue. Each row shows evaluation scores of one metric
for the baseline FCNN, HAYATE, HAYATE-TL and EAZY. The shaded region
in each panel represents the $1\sigma$ range of the individual metric
obtained from EAZY for a given field; the error bars are $1\sigma$. \label{fig:result_fig}}
\end{figure*}

Their performance is evaluated with the metrics for measuring the quality of photo-z point estimates ($\sigma_{\textrm{NMAD}}$ and $\eta_{0.2}$) and output PDFs (KL, CvM and CRPS), as summarised in Table \ref{tab:indicators}.
Each of the metrics is depicted in Fig. \ref{fig:result_fig}, separated by field.
We compare our ML models' performance with EAZY, the underlying template fitting algorithm, whose $1\sigma$ range of the individual metric is represented by the shaded region in each panel.

\S \ref{subsec:HAYATE-v.s.-EAZY} and \ref{subsec:Improvements-with-transfer} describe the perfomance of HAYATE and HAYATE-TL.
In \S \ref{subsec:FCNN-v.s.-CNN}, we discuss the benefit of our simulation-based CNN method, which outperforms the other ML approaches.
\S \ref{subsec:photoz_outliers} presents example archetypes of photo-z outliers useful in exploring the limitations of, and potential improvements to HAYATE when dealing with catastrophic errors in photo-z estimation.

\subsection{HAYATE v.s. EAZY}

\label{subsec:HAYATE-v.s.-EAZY}

\subsubsection{Photo-z statistics}

\label{subsec:Photo-z-statistics}

\begin{figure*}
\begin{centering}
\includegraphics[width=1\textwidth]{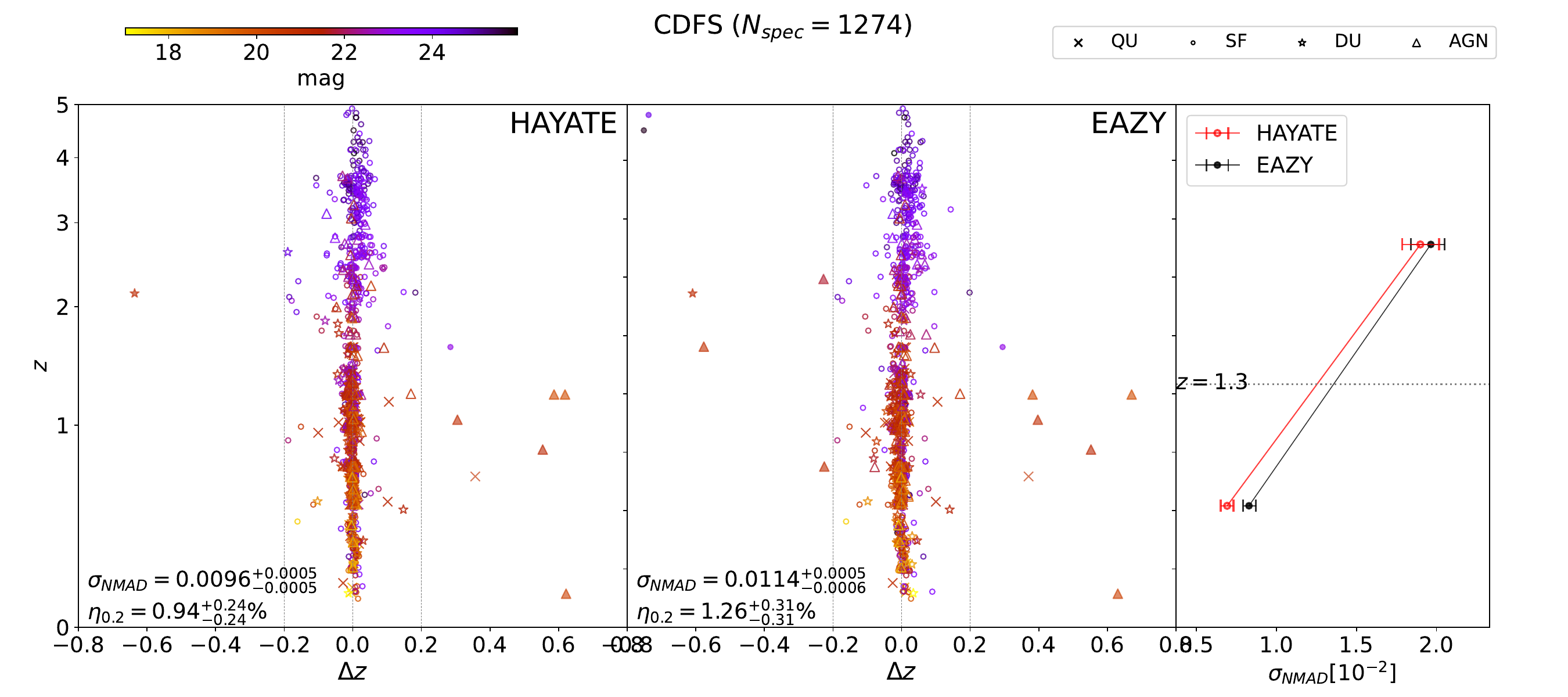}
\par\end{centering}
\begin{centering}
\includegraphics[width=1\textwidth]{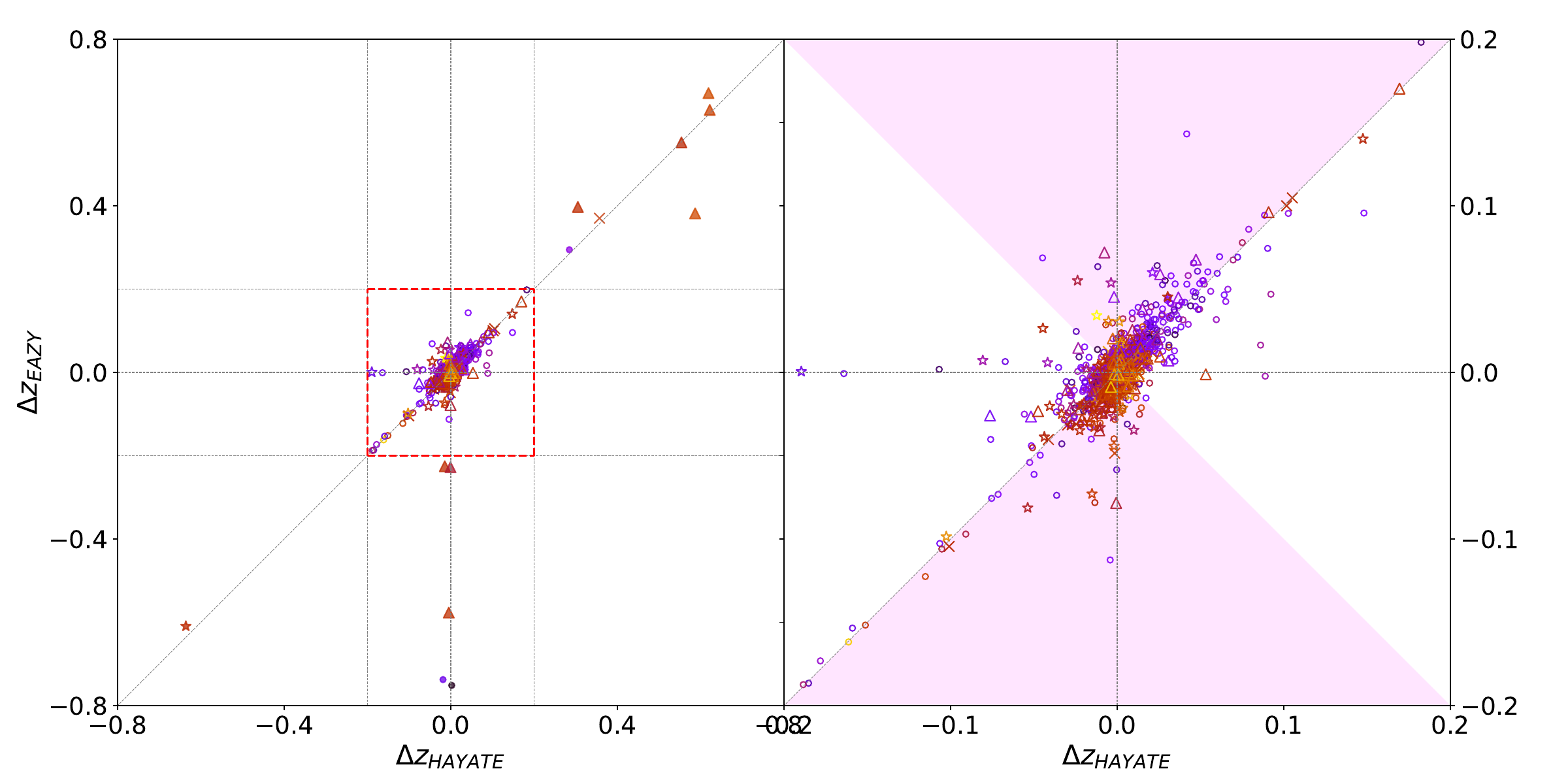}
\par\end{centering}
\caption{Top: distributions of the spec-z catalogue sample for CDFS on the
$z_{\textrm{spec}}-\Delta z$ plane, which are obtained by testing HAYATE (left)
and EAZY (middle). Each data point is presented in different markers
and colours, which represent the galaxy type and $K_{s}$-band magnitude.
The threshold of $\Delta z$ outliers is set to $0.2$, which is shown
by the vertical lines, and the outliers are represented by the filled
markers. The right panel presents the comparison of $\sigma_{\textrm{NMAD}}$
between low-z ($z<1.3$) and high-z ($z>1.3$) samples, individually
derived with HAYATE and EAZY. Bottom: comparison of photo-z errors ($|Z_{ph} - Z_{sp}|$)
between HAYATE ($\Delta z_{\textrm{HAYATE}}$) and EAZY ($\Delta z_{\textrm{EAZY}}$)
for the same sample presented in the top panel. Each panel contains
residual plots of the individual objects, whose entire distribution
is presented in the left panel while the zoom-in plot within the outlier
threshold of $|\Delta z|=0.2$ is in the right panel. The shaded region
represents the area where $z_{\textrm{HAYATE}}$ is better than $z_{\textrm{EAZY}}$.
\label{fig:result_cdfs}}
\end{figure*}

Table \ref{tab:results_tab} compares the performance of HAYATE with EAZY.
We see that HAYATE's point estimates are comparable to, or better than, EAZY, with $\sigma_{\textrm{NMAD}}$ significantly reduced from 1.14 to 0.96 and 1.53 to 1.42 for CDFS and COSMOS, respectively.
Probing the distribution of the test data on a $z_{\textrm{spec}}-\Delta z$ plane further provides insights into how accurate photo-z's can be attained.
The upper row in Fig. \ref{fig:result_cdfs} shows the results for CDFS, where the photo-z's derived with HAYATE ($z_{\textrm{HAYATE}}$) and EAZY ($z_{\textrm{EAZY}}$) are plotted on the left and middle panels, respectively.
Figs. \ref{fig:result_cosmos} and \ref{fig:result_uds} also present the outcomes for COSMOS and UDS (Appendix \ref{sec:Results_for_COSMOS_and_UDS}) in the same manner.
We can see from these figures that the distribution of errors between the two methods are comparable.

We also compare the residuals of HAYATE and EAZY and see a strong correlation.
The bottom row of each figure shows plots of $\Delta z_{\textrm{EAZY}}$
v.s. $\Delta z_{\textrm{HAYATE}}$ for the
test spec-z sample, which represent photo-z errors for EAZY and HAYATE expressed as $\Delta z_{\textrm{EAZY}}=(z_{\textrm{EAZY}}-z_{\textrm{spec}})/(1+z_{\textrm{spec}})$
and $\Delta z_{\textrm{HAYATE}}=(z_{\textrm{HAYATE}}-z_{\textrm{spec}})/(1+z_{\textrm{spec}})$. The data points are generally aligned along 
the diagonal identity line, which demonstrates the ability
of HAYATE to reproduce photo-z estimates that could be provided by
EAZY. The trained network replicates the high accuracy of the template
fitting code but with an execution time $\sim100$ times faster. This indicates
HAYATE can learn to function as a reliable and efficient emulator
of EAZY. 
The resulting mock data should
thus coincide with likely EAZY template fits of the corresponding observed
data.

The results for CDFS and COSMOS show slight improvements in photo-z point estimates with $\sigma_{\textrm{NMAD}}$ reduced by 16\% and 7\%, respectively.
The right panel of each upper row in Fig. \ref{fig:result_cdfs}, \ref{fig:result_cosmos} and \ref{fig:result_uds} shows this is mainly attributable to low-z galaxies, which presents $\sigma_{\textrm{NMAD}}$ separately for sub-samples obtained by splitting the test set with a redshift threshold of $z_{\textrm{spec}}=1.3$.
This can be expected, since the underlying SED templates used for training are constructed with reliable low-z data.
The interpolative nature of ML approaches further surpasses the one-on-one matching of the original template fitting by circumventing the individual template mismatch.

Another significant benefit of our ML method is the ability to generalise to galaxies at higher redshifts in the absence of a large body of high-z data.
The photo-z precision obtained with HAYATE shows no significant difference from that with EAZY even in the extrapolated high-z regime $1.3<z_{\textrm{spec}}<5$.
The extensibility of the target redshift range ensures that the simulations are sufficiently effective beyond the training domain of firm underlying knowledge.

\begin{table*}
\begin{centering}
\caption{Classification of the test objects based on their photo-z's estimated
with HAYATE and EAZY. Objects of Class A are photo-z outliers for
EAZY whose estimates are significantly improved with our ML method
so that they are no longer outliers for HAYATE. Class B contains a
few galaxies on which EAZY conversely outperforms, while photo-z outliers
for both models are classified into Class C. Class D includes `normal'
galaxies whose photo-z's provided by both models are not catastrophically
wrong with respect to the spec-z's. \label{tab:N_outlier}}
\par\end{centering}
\centering{}%
\begin{tabular}{cccccc}
\hline 
\multirow{2}{*}{Object class} & \multirow{2}{*}{Description} & \multirow{2}{*}{Definition} & \multicolumn{3}{c}{Number of objects}\tabularnewline
 &  &  & CDFS & COSMOS & UDS\tabularnewline
\hline 
Class A & Photo-z outliers only for EAZY & $|\Delta z_{\textrm{HAYATE}}|<0.2$, $|\Delta z_{\textrm{EAZY}}|\geq0.2$ & 5 & 6 & 3\tabularnewline
Class B & Photo-z outliers only for HAYATE & $|\Delta z_{\textrm{HAYATE}}|\geq0.2$, $|\Delta z_{\textrm{EAZY}}|<0.2$ & 0 & 1 & 0\tabularnewline
Class C & Photo-z outliers both for HAYATE and EAZY & $|\Delta z_{\textrm{HAYATE}}|\geq0.2$, $|\Delta z_{\textrm{EAZY}}|\geq0.2$ & 11 & 8 & 1\tabularnewline
Class D & Plausible photo-z estimates & $|\Delta z_{\textrm{HAYATE}}|<0.2$, $|\Delta z_{\textrm{EAZY}}|<0.2$ & 1257 & 723 & 308\tabularnewline
\hline 
 &  &  &  &  & \tabularnewline
\end{tabular}
\end{table*}

Applying ML approaches to the template fitting algorithm also produces
a more robust photo-z estimator than EAZY. The outlier rate $\eta_{0.2}$
of estimated photo-z's significantly drops from 1.26\% to 0.94\%,
1.90\% to 1.22\% and 1.28\% to 0.32\% for CDFS, COSMOS and UDS, respectively,
as shown in Table \ref{tab:results_tab}. HAYATE is therefore less
prone to catastrophic failures in photo-z predictions, performing well
on 14 test sources whose photo-z errors would be outliers if derived
with EAZY. We classify them as Class A, along with other sample groups
defined based on a set of $\Delta z_{\textrm{HAYATE}}$ and $\Delta z_{\textrm{EAZY}}$
for each galaxy as presented in Table \ref{tab:N_outlier}. Class
B, on the other hand, only contains 2 galaxies, which are photo-z
outliers for HAYATE but not for EAZY.

20 catalogue objects classified as Class C in Table \ref{tab:N_outlier} show catastrophic solutions for photo-z computation with both models.
Their wrong photo-z estimates are, however, quite similar between the two different methods, which are plotted on the diagonal in each bottom panel of Fig. \ref{fig:result_cdfs}, \ref{fig:result_cosmos} and \ref{fig:result_uds}.
From visual inspection, they are obviously not well-represented by the SED template set of EAZY, which indeed includes some rare objects such as 10 AGNs and 3 dusty star-forming galaxies.
Improving photo-z estimations for Class C objects thus requires extending the population of galaxy templates used by EAZY and for simulating our high-redshift sources. \citet{Brescia+19} also note that to increase the accuracy of AGN photoz's with template-fitting methods, the inclusion of morphological information (extended or point-like) likely provides the biggest improvement.
Example plots of inputs and outputs for these sources are shown in Fig. \ref{fig:agn_outliers}, which are further discussed in $\S$ \ref{subsec:photoz_outliers}.

\subsubsection{PDF statistics}

\label{subsec:PDF-statistics}

\begin{figure*}
\begin{centering}
\includegraphics[width=0.5\textwidth]{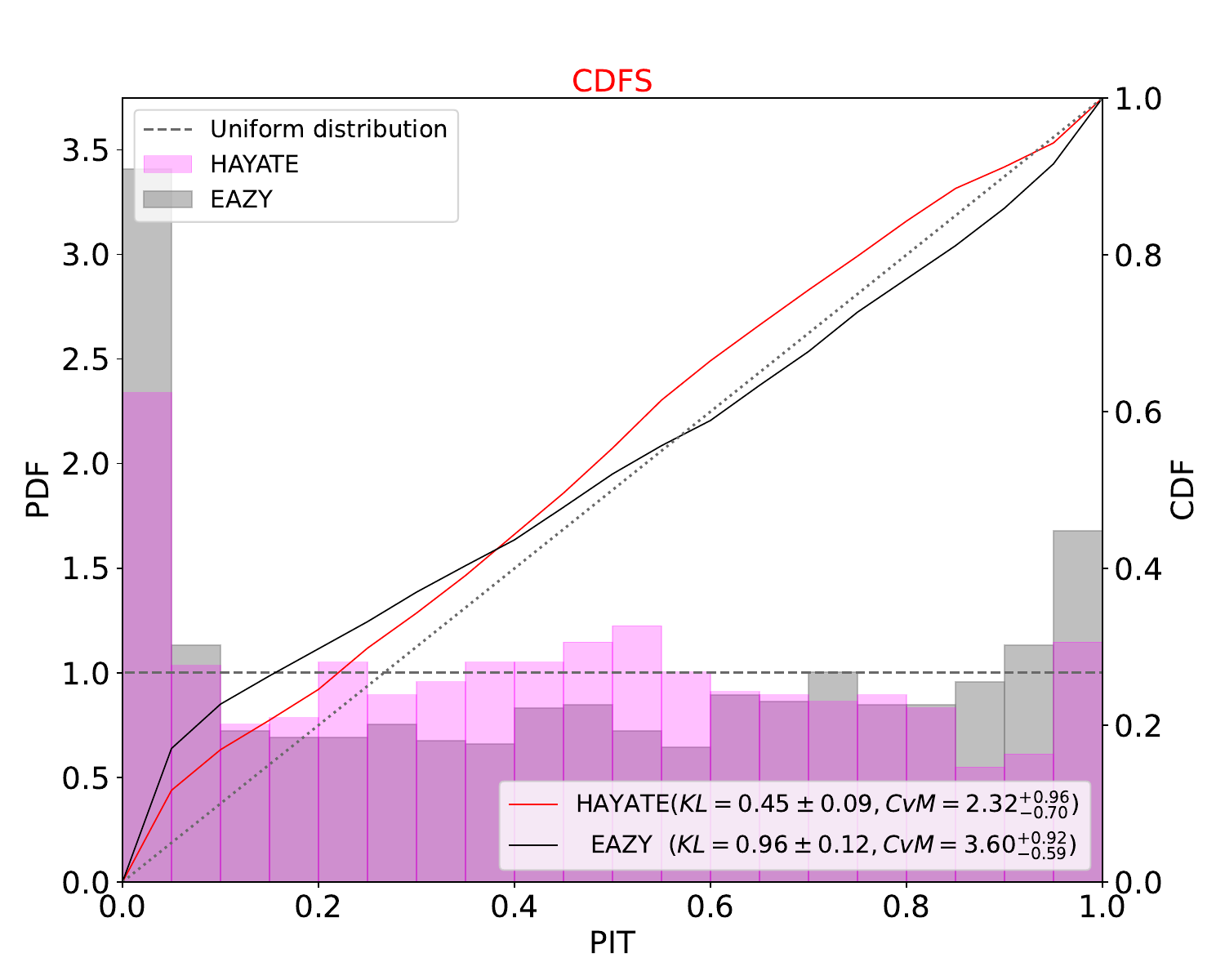}\includegraphics[width=0.5\textwidth]{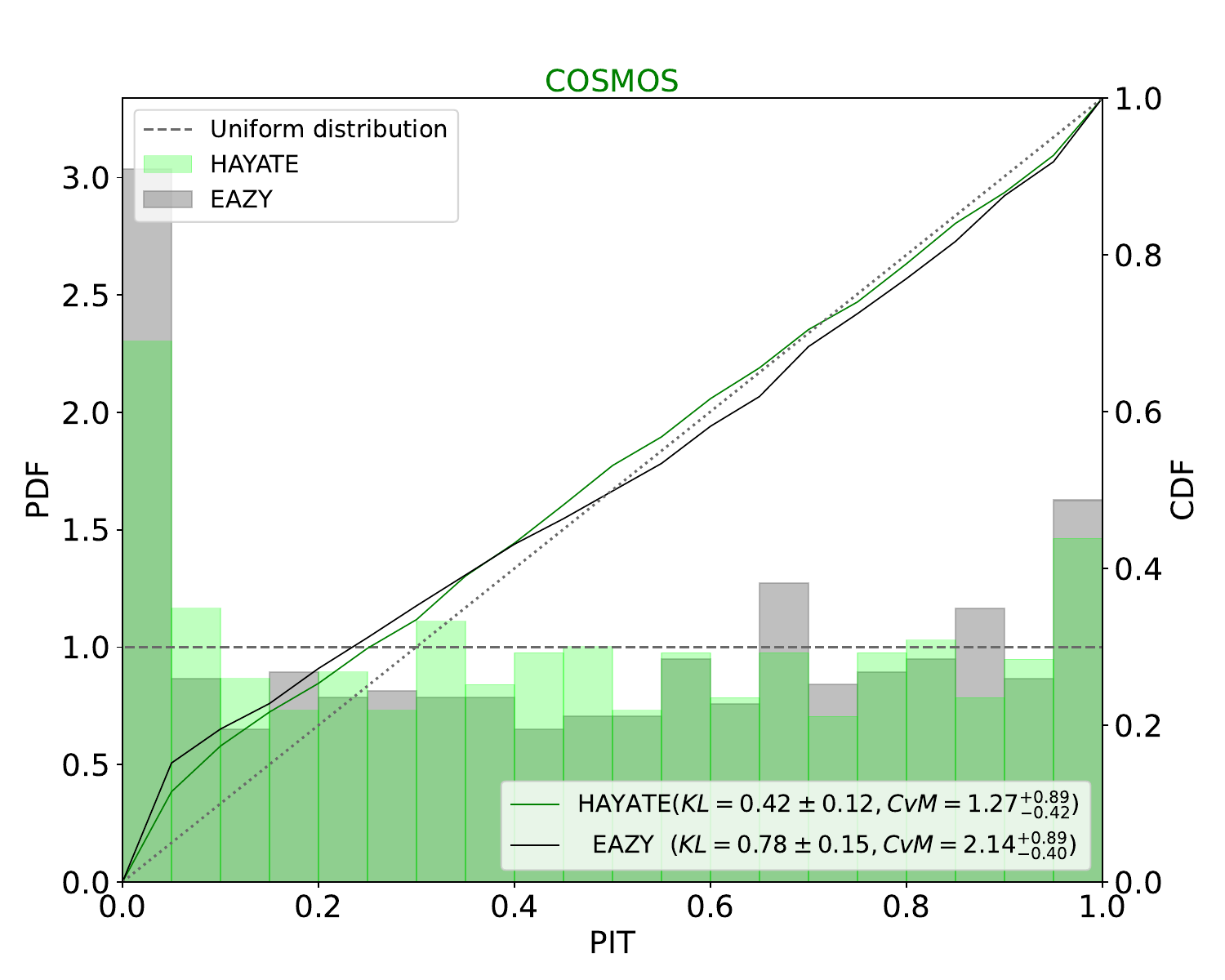}
\par\end{centering}
\caption{PDFs of PIT distributions and their CDFs for the test catalogue samples
in CDFS (left) and COSMOS (right). The PIT distributions derived with
HAYATE and EAZY are shown in the red(CDFS)/green (COSMOS) and gray
bar charts, respectively. The corresponding CDFs are given by the
solid lines in the same colours. The dotted horizontal and diagonal
lines represent the uniform PDF and the corresponding identity CDF,
respectively, indicative of the perfectly calibrated distribution.
\label{fig:pit_dist}}
\end{figure*}

The quality of output PDFs is generally improved with
our ML method as measured by KL, CvM and CRPS, detailed  in Table \ref{tab:results_tab}
and shown in Fig. \ref{fig:result_fig}. HAYATE particularly shows better PIT
distributions for CDFS and COSMOS, with KL and CvM significantly lower
than those derived with EAZY. Fig. \ref{fig:pit_dist} presents the
PIT histograms of HAYATE and EAZY for these two fields, along with
their CDFs used for quantifying deviations from uniformity. KL 
provides a dissimilarity measure between the predictive and uniform
distributions of PIT, while CvM is a CDF-based metric intuitively
represented by the area filled between the corresponding CDF and the
identity line. We can see that the PIT distribution of
HAYATE looks flatter than that of EAZY, which is reflected by the
smaller KL and CvM, indicative of better-calibrated PDFs.

\begin{figure*}
\centering{}\includegraphics[width=0.33\textwidth]{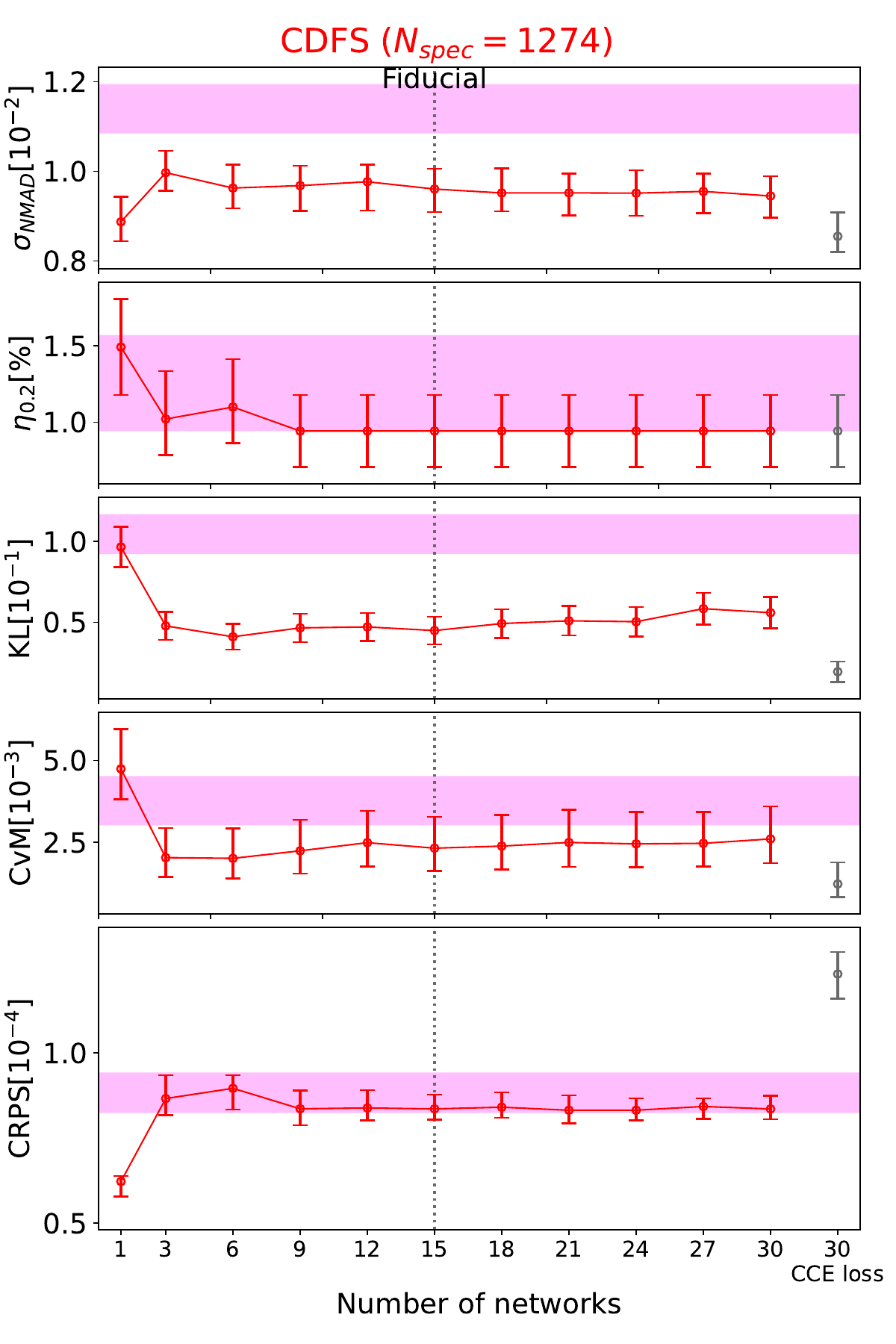}\includegraphics[width=0.33\textwidth]{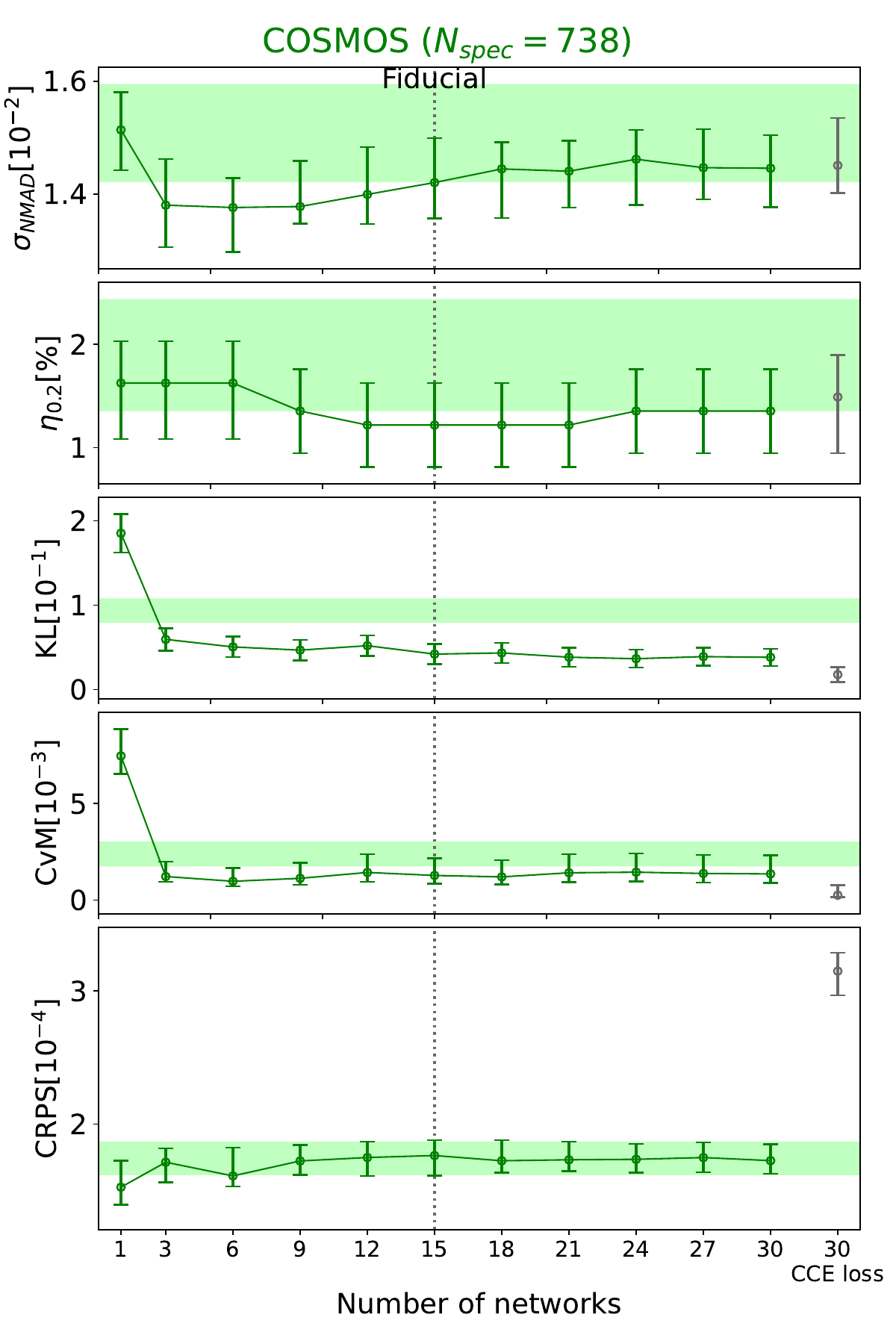}
\includegraphics[width=0.33\textwidth]{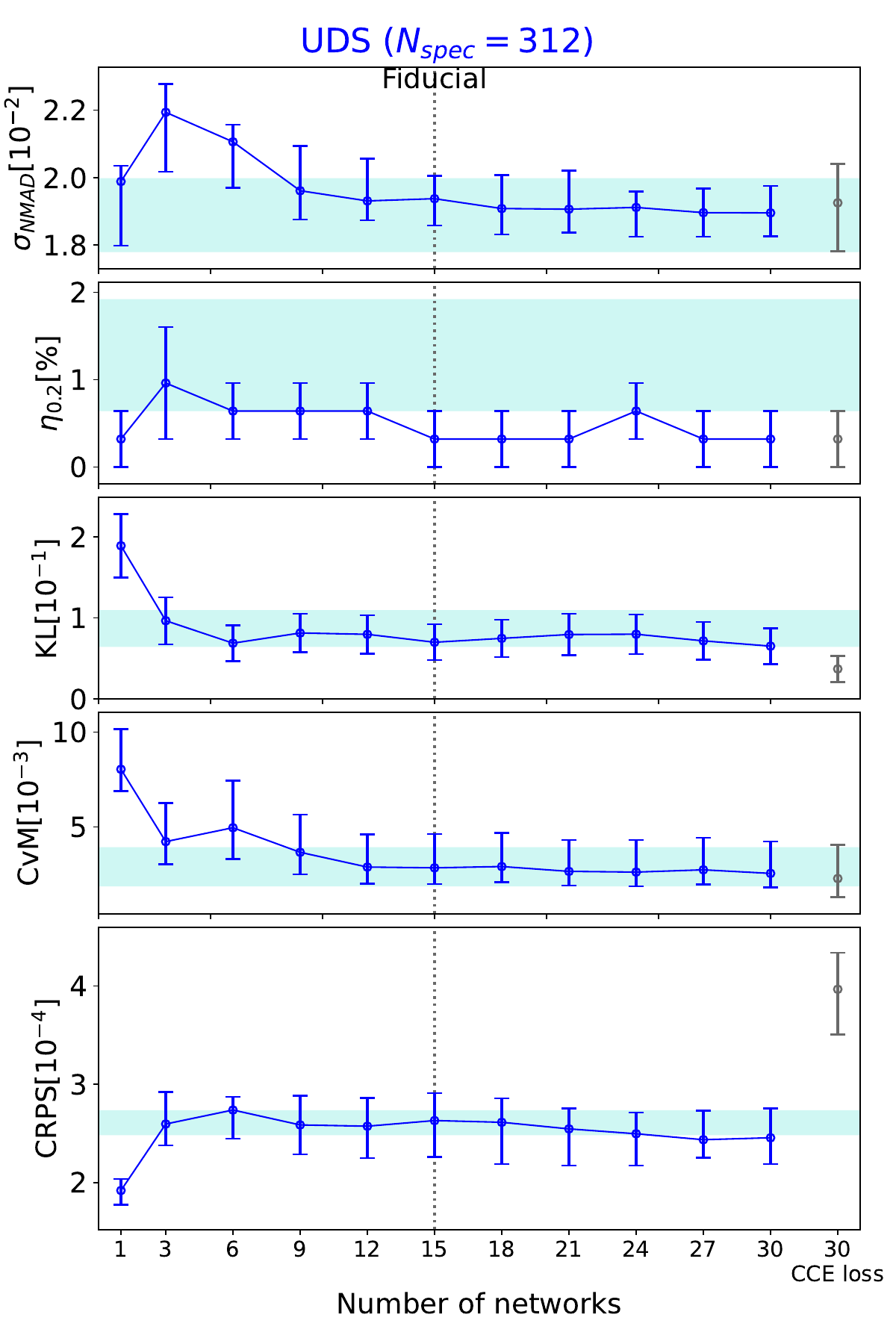}\caption{Changes in photo-z and PDF metrics with the number of CNNs whose predictions
are combined into an ensemble. Each lower-level network is trained
on a bootstrapped sub-sample for a given noise realisation. The fiducial
configuration of HAYATE incorporates 15 components, which is shown
by the vertical dotted lines. The results for CDFS, COSMOS and UDS
are provided in red, green and blue, respectively, while the shaded
regions represent the $1\sigma$ ranges of the corresponding metrics
estimated for EAZY. The left-most column presents the result for a
single CNN that is trained on the entire training sample, while the
last column shows the metrics obtained from the model purely trained
with the CCE loss function without CRPS loss contribution. \label{fig:Nseed}}
\end{figure*}

A major contributor to this is the application of ensemble learning
to generating the combined PDFs. Fig. \ref{fig:Nseed} shows the metrics
estimated for different numbers of CNNs whose individual predictions
are combined into the ensemble PDF. The single network is trained
on the whole training sample, while the multiple models are built
with ensemble learning, as discussed in \S \ref{subsec:Ensemble-learning}.
Increasing the number of networks remarkably improves KL and CvM,
which converges as the number of individual PDFs increases to $\sim 15$.
Our fiducial configuration therefore uses 15 networks, as depicted by the vertical
line in each panel.
ML photo-z codes typically provide PIT distribution characteristics
of convex shape, indicative of overly broad PDFs that are unlikely
to include spec-z's in their tails \citep{Schmidt+20}.
The broadening of PDFs suggests an intrinsic function of ML training approaches
that adds implicit smoothing to the effective error scale \citep{wolfBayesianPhotometricRedshifts2009}.
Conversely, HAYATE produces redshift PDFs whose PIT distribution is rather similar to
that obtained with the template fitting code EAZY, as one can see
in Fig. \ref{fig:pit_dist}. An
over-representation of extreme values is evidenced by a concave histogram, implying
overly narrow PDFs. The ensemble learning technique is aimed at alleviating
this tendency by combining multiple predictions.

\cite{Eriksen+20} have demonstrated a similar effect whereby 
combining multiple networks reduces the number of objects with the lowest
and highest PIT values. The improvement in PIT distribution proved
to be caused by decreasing photo-z outliers. However, this does not
apply to HAYATE, since the outlier rate does not significantly drop
with multiple PDFs combined. Our ensemble approach 
samples the potential solution space 
from different local minima on the loss surface.

We assess the overall form of PIT distributions to probe if the output
PDFs are well calibrated on average. This requires that the CDF value
at the spec-z should be random, rather than that each PDF is well
constrained with respect to its true redshift. The quality of the
individual PDFs thus has to be assessed in conjunction with CRPS,
which represents a distance between the CDF of a given PDF and a step
function with a step at its spec-z. In a derivative form, this can
be translated into how different the PDF is from the corresponding
delta function. We nonetheless often find a trade-off between KL/CvM
and CRPS, or the uniformity of the PIT distribution and the sharpness
of each PDF. A simple explanation for this is that a narrower PDF tends
to be better constrained at its spec-z with lower CRPS, which is more
likely to increase the number of PIT values at extreme edges.

We see in Fig \ref{fig:Nseed} that ensemble learning also improves CRPS
slightly as the number of networks increases. At convergence, 
HAYATE's PDF statistics are each comparable to, or better than, EAZY.
The validity of each output
PDF is further supported by the sufficiently small CRPS. In Fig.
\ref{fig:Nseed}, we see that the primary contribution to this is
incorporating the CRPS term into the joint loss function. The results
for the model trained exclusively with the CCE loss are presented
on the rightmost columns in each panel. The inclusion of CRPS in the
training loss function remarkably decreases the CRPS of output PDFs. 

The analysis on the PDF statistics indicates the output
PDFs derived with HAYATE are more reliable than those obtained with
EAZY, with respect to spec-z's, while their overall population is
statistically more self-consistent. We thus demonstrate HAYATE attains
good-quality photo-z PDFs by leveraging the benefits of multiple approaches,
which performs as an ensemble of lower-level networks optimised for
the joint loss.

\subsection{Improvements with transfer learning}

\label{subsec:Improvements-with-transfer}

Fig. \ref{fig:result_fig} also demonstrates further improvements 
in photo-z precision due to transfer learning. The HAYATE-TL model
shows an improved precision for all  fields, with reduction in $\sigma_{\textrm{NMAD}}$
by $30\%$ (CDFS), $13\%$ (COSMOS) and $7\%$ (UDS), respectively.
By leveraging empirical data,  HAYATE-TL
achieves more accurate photo-z estimates than HAYATE and EAZY with
no more than $\lesssim1000$ additional observational training examples.
The enhancement of the model's performance nonetheless correlates with the sample size
of spec-z data available. Accomplishing better performance with transfer
learning is likely with a larger sample size.

The higher precision of photo-z point estimates indeed results from
improved output PDFs yielded by HAYATE-TL. Table \ref{tab:results_tab}
shows significantly better CRPS for CDFS and COSMOS, which reduces
from 0.83 to 0.70 and 1.76 to 1.48 with the assistance of spec-z information
for re-training. The fine-tuning of the pre-trained network thus better
constrains each redshift PDF over the peak around the spec-z. The improved PDF consequently provides a more precise photo-z point estimate.

The HAYATE-TL photo-z outliers still comprise exactly the same objects
as HAYATE. The outlier rate consequently remains
the same despite the re-training with observations.
This reflects the limitation of transfer learning from simulations,
which exclusively benefits an `ordinary' test object whose colour-redshift
relation can be fine-tuned by training with the remaining sample.
It can not adapt to `anomalous' sources whose photometric
data are not sufficiently represented in the training set
along with reliable spec-z's.
We therefore conclude the photo-z outliers found with HAYATE are
intrinsically inconsistent with the input-output mapping derived 
from both the template-based training set and the observed data.

Table \ref{tab:results_tab} also presents the result for the CNN
model which is trained on the same spec-z samples as used for transfer
learning but completely from scratch. HAYATE-TL significantly outperforms
the most common method of training purely with spec-z data, although
both fundamentally learn with the same observed samples. Training
with simulations proves to supplement the insufficient spec-z sample
for training. These results indicate transfer learning is effective
for making a minor adjustment both for output redshift PDFs and their
single-point estimates using spec-z information. 
Training with simulations lays important groundwork for the subsequent observational fine-tuning. 

\subsection{CNN v.s. FCNN}

\label{subsec:FCNN-v.s.-CNN}

Evaluating the metrics for the two ML models, the baseline FCNN and
HAYATE, reveals the CNN-based architecture shows overall improvements
on photo-z point estimates. Table \ref{tab:results_tab} and Fig.
\ref{fig:result_fig} show significant drops in $\sigma_{\textrm{NMAD}}$,
$\eta_{0.2}$ and CRPS for all the fields. This indicates that the CNN is more
likely to yield reliable redshift PDFs with precise point estimates than a FCNN.

The superior performance of HAYATE compared to the FCNN model indicates the benefit
of prioritising local feature extraction from the combined arrays
of fluxes and photometric errors. CNNs are particularly suited to
high-dimensional data such as image processing since convolutional kernels require
many fewer trainable parameters than FCNNs. Convolution operations
are performed primarily for extracting local information and preserving
the spatial relationship between features; these features become more
abstract from layer to layer through the network. 
The CNN photo-z models have commonly been
trained on galaxy images instead of photometry summary information, which allows for learning
with supplementary information on the spatial flux distribution \citep{Pasquet+19,Schuldt+21,Henghes+22,Lin+22,Zhou+22}.
Our ML instead leverages the demonstrated ability of CNNs to 
capture and interpret the local features of galaxy
SEDs obtained from the flux distribution over a range of wavelengths.

\subsection{Analysis of individual redshift PDFs}

\label{subsec:photoz_outliers}

\begin{figure*}
\begin{centering}
\includegraphics[width=1.\textwidth]{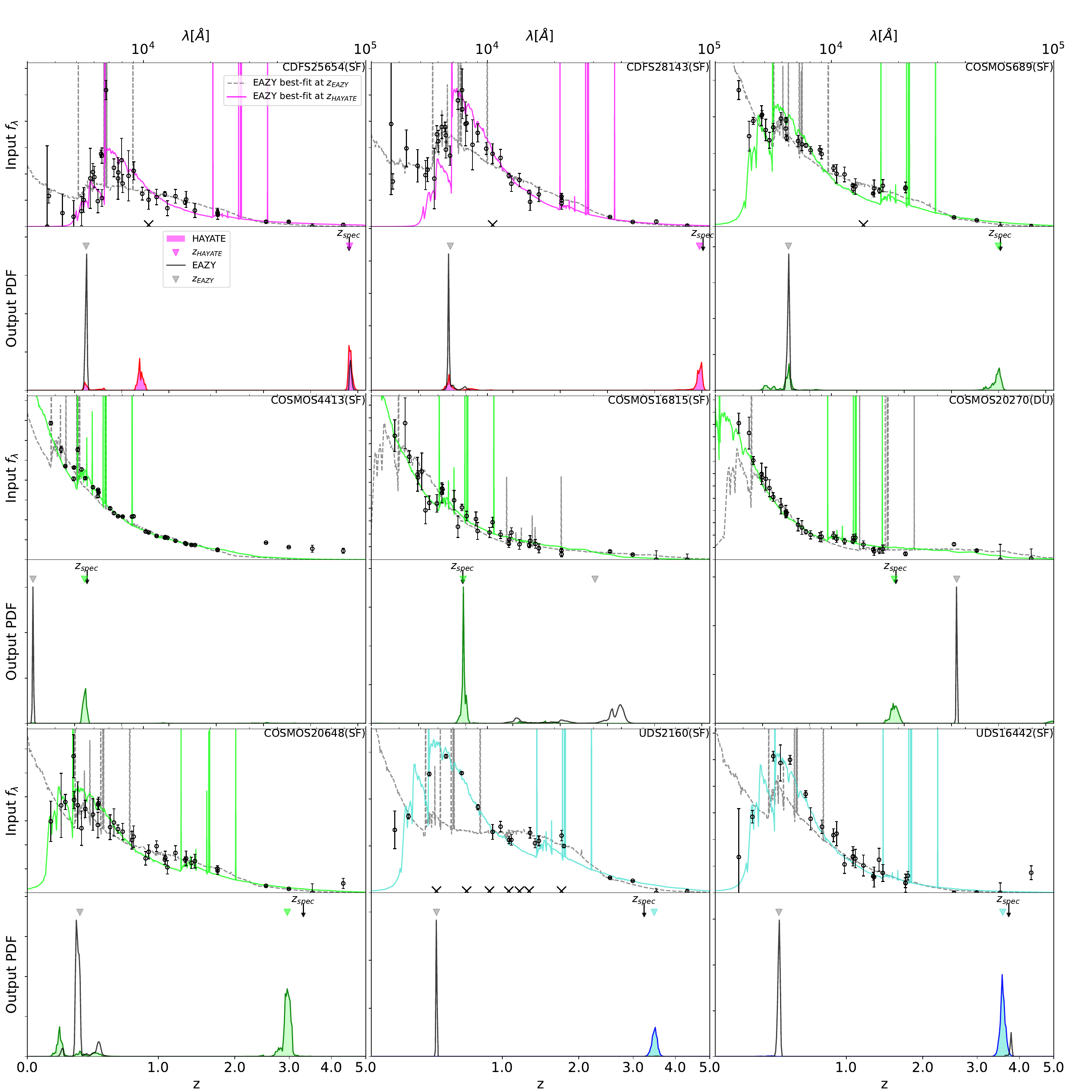}
\par\end{centering}
\caption{Example star-forming galaxies of Class A. The input fluxes with errors
are plotted in the top row of each panel, along with the missing values
represented by the black crosses. The gray dotted line show the best-fit
SEDs optimised with the photo-z by EAZY ($z_{\textrm{EAZY}}$), while the coloured
solid line represents the result at the fixed photo-z derived
with HAYATE ($z_{\textrm{HAYATE}}$). The bottom row compares the photo-z PDFs
produced by HAYATE and EAZY, shown by the shaded region and the solid
black line, respectively. For each object their photo-z point estimates are given
by the coloured and gray upside-down triangles, while the spec-z by the down arrow. \label{fig:sf_outliers_for_eazy}}
\end{figure*}

The robustness of our method can be explored with by visual inspection
of individual PDFs predicted by HAYATE and EAZY. We particularly focus
on the test objects classified as Class A, defined in Table \ref{tab:N_outlier},
which are responsible for the improved outlier rate vs template fitting approaches.
Fig. \ref{fig:sf_outliers_for_eazy} shows example star-forming
galaxies of Class A, whose input photometric data and output PDFs
are presented in the top and bottom rows of each panel, respectively.
HAYATE obviously performs better than EAZY on these objects, providing
more reliable PDFs with respect to their spec-z's, which are represented
by the black circles on the horizontal axes. This results in
more accurate photo-z point estimates than the EAZY predictions, shown
by the coloured and gray vertical lines.

We can gain further insight into the reason for the improvements by
probing the best-fit SEDs derived with EAZY when fixed to the photo-z's.
They are represented by the coloured lines plotted with the input
fluxes, which can be compared to the gray dotted lines for the corresponding
SEDs of EAZY photo-z's. One major failure of template fitting is to misinterpret
the spectral features of the Lyman and Balmer breaks, or the Lyman-alpha
and Balmer emission lines \citep{Benitez00, Brammer+08}. Some of the PDFs produced by HAYATE indeed
show minor peaks around their corresponding EAZY photo-z's, 
showing the learned degeneracy inherited from the original template fitting algorithm.

\begin{figure*}
\begin{centering}
\includegraphics[width=1.\textwidth]{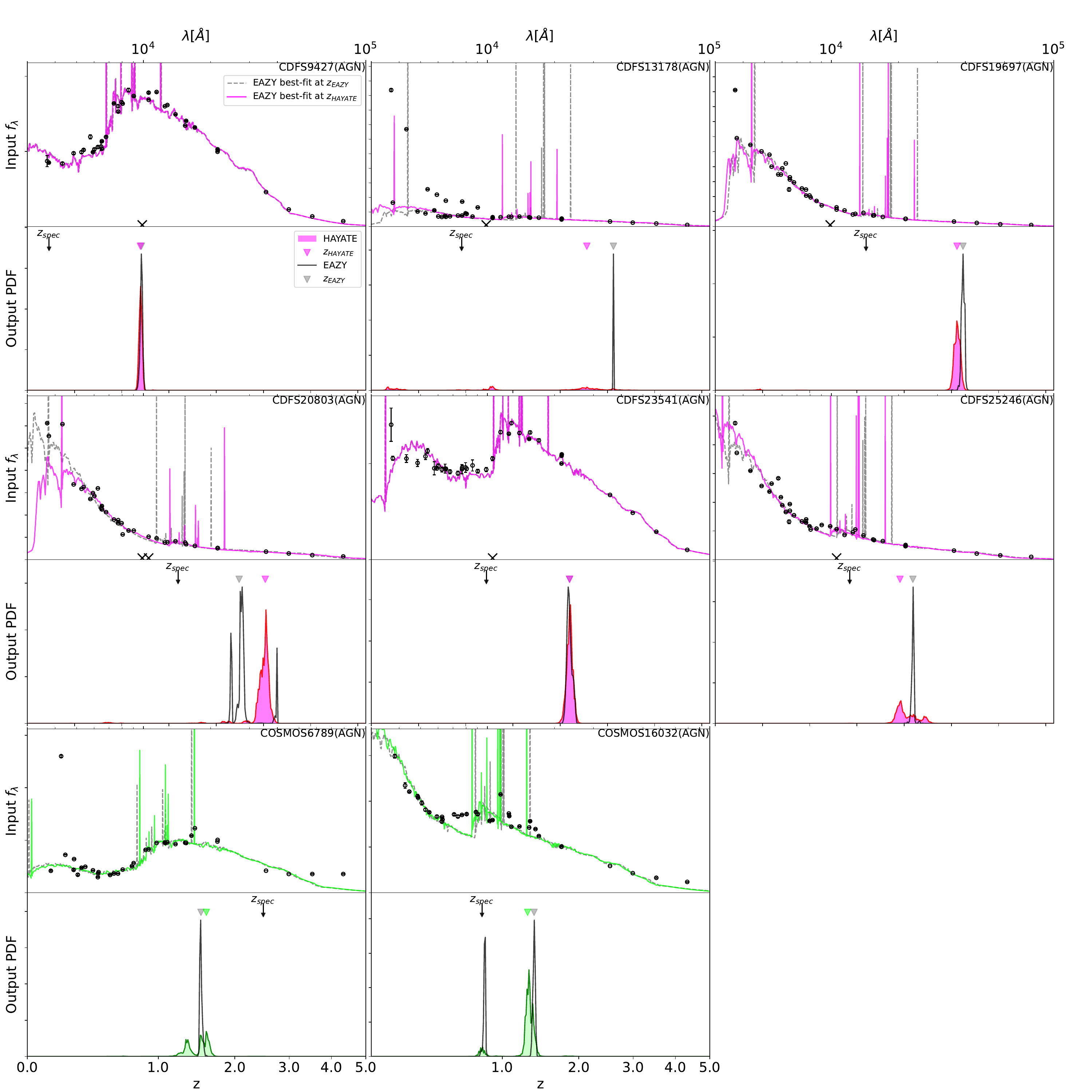}
\par\end{centering}
\caption{Example AGNs of Class C in CDFS and COSMOS. \label{fig:agn_outliers}}
\end{figure*}

We may glean further clues for improving HAYATE's output
PDFs by investigating typical photo-z outliers, although the outlier
rate is low with $\eta_{0.2}\lesssim1\%$. One major
group consists of rare objects whose photometric data are not well
represented by the training samples. We can see some likely AGN data points
(estimated from visual inspection of spectra)
in the lower left panels of Fig.\ref{fig:result_cdfs} and \ref{fig:result_cosmos},
depicted by triangles outside the region of $|\Delta z_{\textrm{HAYATE}}|<0.2$
and $|\Delta z_{\textrm{EAZY}}|<0.2$. These objects are included in Class
C, whose photo-z's are outliers both for HAYATE and EAZY often with
similar point estimates. For the example AGNs presented in Fig. \ref{fig:agn_outliers},
both models provide PDFs of erroneous photo-z's, although their input
photometric data are obtained with the brightest magnitudes and the
highest SNRs. Most of them show no minor peaks at the spec-z's in
the distributions. This reveals the ensemble of standard EAZY templates
can not intrinsically cover galaxy SEDs of some rare objects, which
means that the simulated training datasets will also lack objects
of this class, and such anomalous objects result in catastrophically wrong solutions.

\begin{figure*}
\begin{centering}
\includegraphics[width=1.\textwidth]{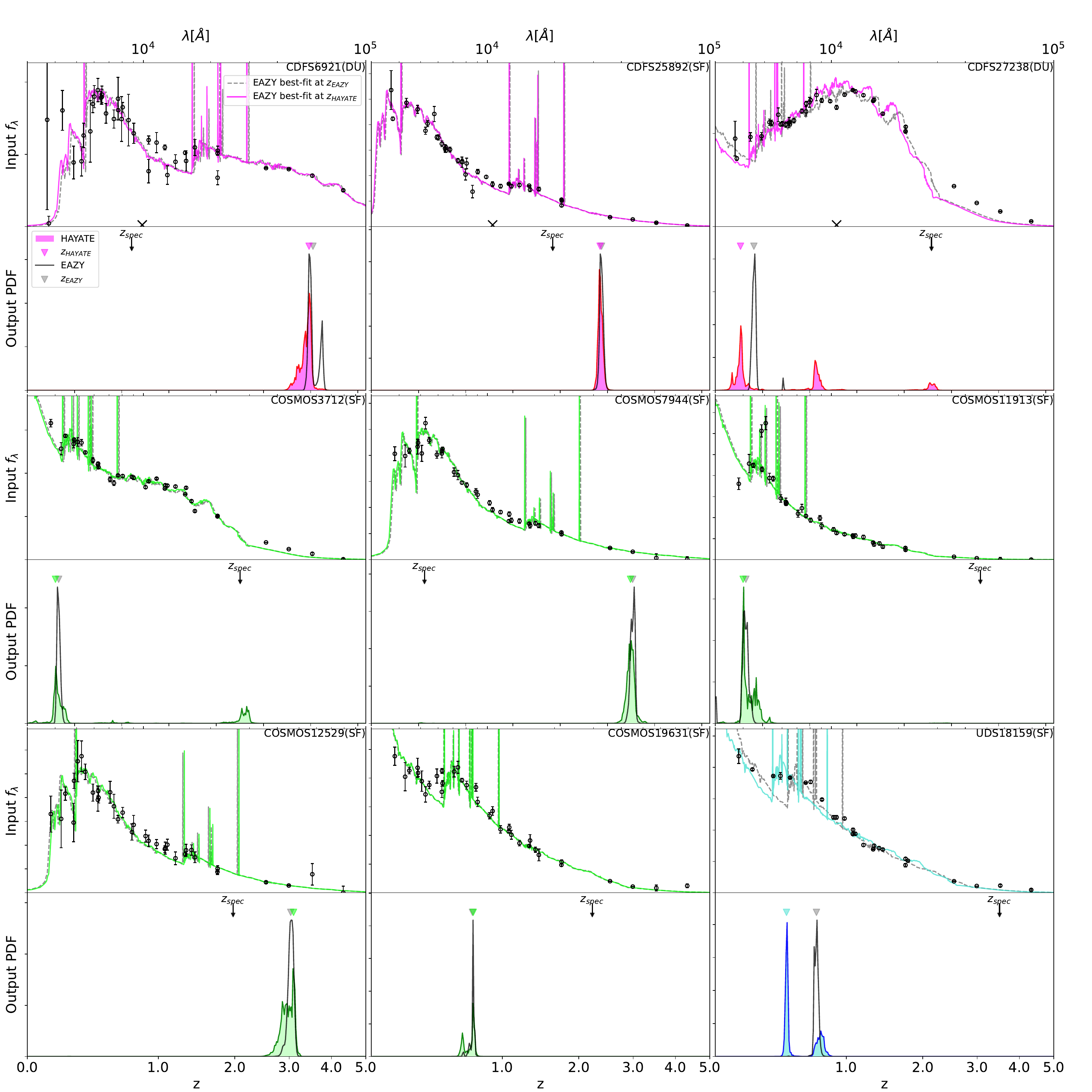}
\par\end{centering}
\caption{Example star-forming galaxies of Class C. \label{fig:sf_outliers_for_both}}
\end{figure*}

Class C also contains some star-forming galaxies which result in
incorrect photo-z predictions. They
are varied in photometric SNR and derived spec-z's,
whose results are presented in Fig. \ref{fig:sf_outliers_for_both}.
HAYATE and EAZY both predict quite similar photo-z's for each object,
although they are significantly divergent from the spec-z. One
can find the best-fit SEDs derived at $z_{\textrm{HAYATE}}$ and $z_{\textrm{EAZY}}$
which indeed look well fitted to the input fluxes. 
A deficiency in the template set and thus HAYATE's training data
means that neither can correctly classify these difficult objects.

Photo-z point estimates of HAYATE are generally correlated with those
computed by EAZY, as discussed in $\S$ \ref{subsec:Photo-z-statistics}.
The network is clearly able to exploit the demonstrated ability of 
template fitting to predict precise photo-z's.
Assessing the
model's performance on the individual outliers further demonstrates
how our hybrid approach has internalised the fundamental functions of
EAZY, including even the failure to produce reliable PDFs for some
difficult objects. One remarkable benefit of HAYATE is, however,
the potential to remedy the vulnerability of EAZY to the misinterpretation
of spectral features, particularly characterised by the Lyman and Balmer breaks.

\section{Discussion}
\label{sec:Discussion}

We have demonstrated the potential for HAYATE to contribute to efficient data mining for future large surveys with the following benefits:
\begin{enumerate}
\item Our method can be applied to a broad redshift range including high-z galaxies which are deficient in reliable observational data for training. The network trained with template SEDs from EAZY can function as a reliable emulator,
with $\sim100$ times shorter running time.
\item The analysis of $\sigma_{\textrm{NMAD}}$ reveals that in the interpolative regions of the low-z colour space, the ML methodology results in higher accuracy in photo-z estimation than the original template fitting approach. It also performs comparably well even in the high-z extrapolative regime.
\item HAYATE is likewise more robust to photo-z outliers than EAZY. In particular, its output photo-z PDFs are less vulnerable to the degeneracy of redshift caused by misinterpretation of the Lyman and Balmer breaks from input photometric data.
\item Optimising the joint loss comprising $L_{\textrm{CCE}}$ and $L_{\textrm{CRPS}}$ keeps the credibility of individual PDFs comparable to that for EAZY in terms of CRPS.
\item Ensemble learning shows significant improvements in KL and CvM, which enables HAYATE to provide redshift PDFs better calibrated than EAZY with a flatter PIT distribution.
\item Transfer learning with HAYATE-TL significantly improves the model's performance further, achieving more reliable photo-z point estimates and PDFs than HAYATE. This reduces $\sigma_{\textrm{NMAD}}$ by $\sim5-30\%$ depending on the sample size of spec-z datasets. We expect to benefit from the fine-tuning with spec-z information for future photo-z studies since spectroscopy will be conducted along with imaging in many upcoming survey projects.
\item Training with simulations shows remarkable improvements in both photo-z and PDF statistics compared to the purely observation-based training. This enables us to utilise ML techniques for redshift estimations where only small spec-z samples are available: in this work, no more than 1274, 738 and 312 objects in CDFS, COSMOS and UDS, respectively.
\item Our empirical noise application method allows any missing values to be included in the input photometric data. This can enhances photometric resolution and spectral coverage of the target photometric sample without reducing sample size, which is compiled from multiple sub-catalogues by cross-matching between individual sources of different many-band photometry.
\end{enumerate}

It is worth pointing out that, although a hybrid method optimised to perform well in the high-redshift regime, there is no clear step-change in performance beyond a certain threshold. HAYATE performs well across the entire parameter range, exploiting the strengths of the two approaches previously considered somewhat disjoint.

Exploring billions of objects catalogued by the Stage IV dark energy surveys will require the exploitation of photo-z's at the expense of reliable spec-z information.
A simplistic extrapolation from past campaigns would estimate over 10,000 years of continuous integration time for obtaining spectra of the LSST "gold sample" galaxies \citep{Newman_Gruen22}.
We may need $\gtrsim$ 30,000 spectra in training and calibration of photo-z's for a Stage IV survey, from $\gtrsim 15$ widely-separated fields of $\sim 0.09\;\mathrm{deg^2}$ each;
for instance, this corresponds to an estimated survey time of over a few years even with the Extremely Large Telescope for the LSST survey depths \citep{Newman+15}.
The limited redshift range targeted by the Euclid survey could rather demand a smaller spec-z sample size to meet the cosmology requirements for photo-z calibration, but still exceeding $\sim 5000$ \citep{Stanford+21}.
Efficient and accurate estimation of photo-z\textquoteright s will have fundamental importance in various fields of extragalactic astronomy and cosmology as the pace of follow-up spectroscopy is never able to sufficiently meet the required volume of objects from such imaging surveys.


Template-fitting methods perform well on the current generation of surveys, though at the data volumes of the levels expected for the Stage III and IV surveys the compute power needed becomes a non-trivial issue. A photo-z estimator that is orders of magnitude faster while preserving excellent performance would assist in scaling these data pipelines.

This work builds on the performance of EAZY, applied to S16 (including the data products from the ZFOURGE survey), which covers 128 $\mathrm{arcmin}^{2}$ to a limit of $AB\sim26$ in the $K_{s}$ band for CDFS, producing imaging of $\sim30,000$ galaxies.
The ongoing and future surveys will probe much fainter objects in wider survey areas, producing observational catalogues of unprecedentedly large sample size.

Our hybrid method can be applied to any photometric catalogues by simulating photometry using the corresponding transmission curves for mock SEDs with simulated noise based on the observational errors.
The simulation-based catalogue construction also allows training methods to be extrapolated outside their initial redshift ranges, i.e. $z<1.3$ in this work.
We set the upper bound of the target redshift range to 5 considering the number of spec-z data available as the test samples.
This can be reasonably extended to much higher redshifts depending on the target survey or photometric catalogue.
Some recent works have proposed using simulated photometric data in training photo-z networks \citep[e.g.,][]{Eriksen+20,Ramachandra+22}, but the redshift range still covers up to $\sim1.2$ at the highest.

The mainstream in traditional ML approaches have involved training a photo-z algorithm exclusively with spec-z information \citep[e.g.,][]{Firth+03,Brescia+13, Brescia+14,Bonnett15,Sadeh+16,Jones_Singal17}.
The accuracy of predicted photo-z\textquoteright s essentially depends on the quality and completeness of the training dataset, which requires large spec-z samples.
The target redshift range for ML has been thus limited to low-z regions of sufficient spectroscopic completeness.
This accounts for the current prevalence of template-based methods for high-z galaxies, although ML approaches are rather common at $z\lesssim1$.
The extensibility of the target redshift range is one critical functionality of our hybrid method that will enable to infer accurate photo-z's of faint high-z galaxies obtained from the upcoming survey projects.

We have also demonstrated the potential of transfer learning to fine-tune the pre-trained model with spec-z information and improve the photo-z precision of normal sources, whose estimations are not outliers with respect to $\Delta z$.
This will significantly benefit future ML photo-z studies, since spectroscopy will be conducted along with imaging in many upcoming survey projects.

Forthcoming programs from the JWST will provide opportunities to both improve the algorithm and to put it into practice. Deep spectroscopic data will bolster the training set available for transfer learning. For instance, the ongoing JWST Guaranteed Time Observations (GTO) program, the JWST Advanced Extragalactic Survey \citep[JADES;][]{Rieke20,Bunker+20} will provide Near InfraRed Spectrograph (NIRSpec) spectroscopy covering in \textquoteleft DEEP\textquoteright{} survey mode a smaller survey area of 46 $\mathrm{arcmin}^{2}$ in HUDF/GOODS-S but to a much fainter limit of $AB\sim30$.
Another survey mode, the \textquoteleft MEDIUM\textquoteright{} survey, will cover no less than 190 $\mathrm{arcmin}^{2}$ in GOODS-S and GOODS-N to a limit of $AB\sim29$. JADES will observe $\sim5000$ galaxies at $1<z<5$, $\sim2000-4000$ galaxies at $z>5$ and $\sim300$ galaxies at $z>6$.

On the other hand, deep imaging surveys, such as 
Public Release IMaging for Extragalactic Research \citep[PRIMER][]{Dunlop+21} will probe an even larger area of $\sim400$ $\mathrm{arcmin}^{2}$ than GTO in COSMOS and UDS to a limit of $AB\sim28.5$, revealing $\sim100-200K$ galaxies out to $z\sim12$.
The COSMOS-Webb \citep{Kartaltepe+21} will also produce wide area imaging covering 0.6 $deg^{2}$ in COSMOS to a limit of $AB\sim28$, expected to offer near-IR imaging of half a million galaxies along with 32,000 in the mid-IR and identify hundreds of massive quiescent galaxies in the first 2 Gyr ($z>4$).
These imaging datasets will be excellent candidates for applying HAYATE, and via transfer learning will leverage the smaller spectroscopic surveys.

The non-Gaussianity of predicted PDFs also distinguishes HAYATE from other commonly used approaches, which tend to assume the underlying components to be Gaussian \citep{DIsanto_Polsterer18,Eriksen+20,Lima+22}.
HAYATE yields non-parametric PDFs as the outputs of the softmax activation.
Realistic redshift PDFs should indeed contain non-Gaussian properties such as asymmetry and tails, reflecting an interplay of various features in target photometric data:
for instance, filter functions, the set of filters used and their observational error distributions.
They may also be vulnerable to colour-redshift degeneracies, which are represented by multiple peaks.
These individual features could have a significant impact on cosmological measurements \citep{Mandelbaum+08,Palmese+20}.

Nevertheless, HAYATE still fails to return a uniform PIT distribution,
required for applying the the output PDFs to estimating $N(z)$ of an ensemble of galaxies \citep{Newman_Gruen22}.
EAZY is not vulnerable to systematic broadening or narrowing of output PDFs in general \citep{Wittman+16,Schmidt+20}.
The better calibration of redshift PDFs offered by HAYATE thus provides insight into obtaining an even flatter PIT distribution that could meet the requirements for many high-precision cosmology measurements.
Improving the ensemble learning approach, combined with transfer learning depending on the science case, is an potential avenue of research, which has contributed to significantly reducing KL and CvM.

Another issue to be addressed is the fidelity of simulated training data.
The success of transfer learning indicates that there remains an intrinsic disparity in data quality between simulated and observed datasets, and that the mock photometric data used for training can be further improved.
In essence, the quality of the mock SEDs relies on the performance of EAZY, while the noise model affects simulated photometry for a given test sample.
These aspects of our hybrid method simultaneously lead to HAYATE's ability to emulate EAZY and its limitations, while surpassing the performance of the original template fitting code.

\section{Conclusion}
\label{sec:Conclusion}

We have developed a novel photo-z CNN, dubbed HAYATE, based on a hybrid method that incorporates the demonstrated ability of template fitting into the latest empirical modelling.
It is primarily aimed at combing the benefits of ML- and template-based approaches by performing as an efficient ML emulator of EAZY beyond the limitations of spec-z completeness at low-z.
This was achieved by extrapolating the SED coverage obtained from low-z photometric data to higher redshifts.
Technically, we artificially redshifted EAZY best-fit SEDs for the S16 sources of $z<1.3$ such that the training set of mock SEDs covers a broader redshift range up to $z=5$.
Further advancements were likewise explored via simultaneous optimisation of output photo-z PDFs and point estimates, aided by the modern ML techniques: training with the joint loss function (\S \ref{subsec:Training-process}), ensemble learning (\S \ref{subsec:Ensemble-learning}), and transfer learning (\ref{subsec:Transfer-learning}).
The photo-z networks of different configurations, as well as EAZY, were tested on the updated S16 spec-z samples, evaluated based on commonly used performance metrics for measuring the quality of photo-z point estimates and output PDFs: $\sigma_{\textrm{NMAD}}$, $\eta_{0.2}$, KL, CvM, and CRPS, as described in \S \ref{subsec:Indicators-for-photo-z-estimates}.

Considering the applicability of our methodology to a variety of catalogues, HAYATE should generalise to a flexible set of photometric bands.
The current framework is a bespoke solution for a specific catalogue with a fixed combination of broad-band filters.
We may develop an extended architecture where the input involves a broader range of photometric band filters by allowing missing data to be incorporated into those unavailable to a given catalogue.
A single model could then adapt to different catalogues simultaneously by learning on a collection of individual training samples.
An upgraded model is under development and will be the subject of a future work.

Further improvements require a strategy to extend the training sample beyond the scope of EAZY predictions.
The simplest approach would be to incorporate a broader range of galaxy SEDs from external sources into the training set, 
enhancing the model's robustness to those photo-z outliers whose typical SEDs are not included in the EAZY outputs.
This particularly applies to some of the example galaxies discussed in \S \ref{subsec:photoz_outliers} including AGNs.

Blended spectra are a likely source of photo-z errors, and difficult to eliminate in the preprocessing stage. All photo-z methodologies are vulnerable to this source of contamination, for which a correct redshift is not defined. It is possible that future methods, such as ML-based algorithms which directly consume the 2D spectra, could mitigate this further.

Our hybrid method may both benefit from and complement other recent developments. \cite{wangInferringMoreLess2023} have demonstrated promising results by using carefully chosen priors to break the age-mass-redshift degeneracy, and have recovered accurate photo-zs using the Prospector--$\alpha$ \citep{lejaDerivingPhysicalProperties2017} stellar population properties inference code, simultaneously recovering redshift with other stellar properties using Bayesian inference. \citep{wangSBIFlexibleUltrafast2023} exploit simulation based inference \citep[SBI;][]{cranmerFrontierSimulationbasedInference2020}, which allows efficient sampling of computationally-expensive models, to massively accelerate this multi-parameter fitting compared to nested sampling by up to a factor of $10^4$. These methods, applied to simulated JWST data, efficiently recovered photo-zs with comparable outlier rates ($\sigma_{\textrm{NMAD}} \sim 0.04 $) along with multi-modal PDFs.

\section*{Acknowledgments}
KG and CJ acknowledge support from Australian Research Council Laureate Fellowship FL180100060.
Part of this work was performed on the OzSTAR facility at Swinburne University of Technology.

\section*{Data Availability}
The datasets were derived from sources in the public domain at https://zfourge.tamu.edu.

\bibliographystyle{aasjournal}
\bibliography{astro}

\appendix

\section{Results for COSMOS and UDS}
\label{sec:Results_for_COSMOS_and_UDS}

\begin{figure*}
\begin{centering}
\includegraphics[width=1\textwidth]{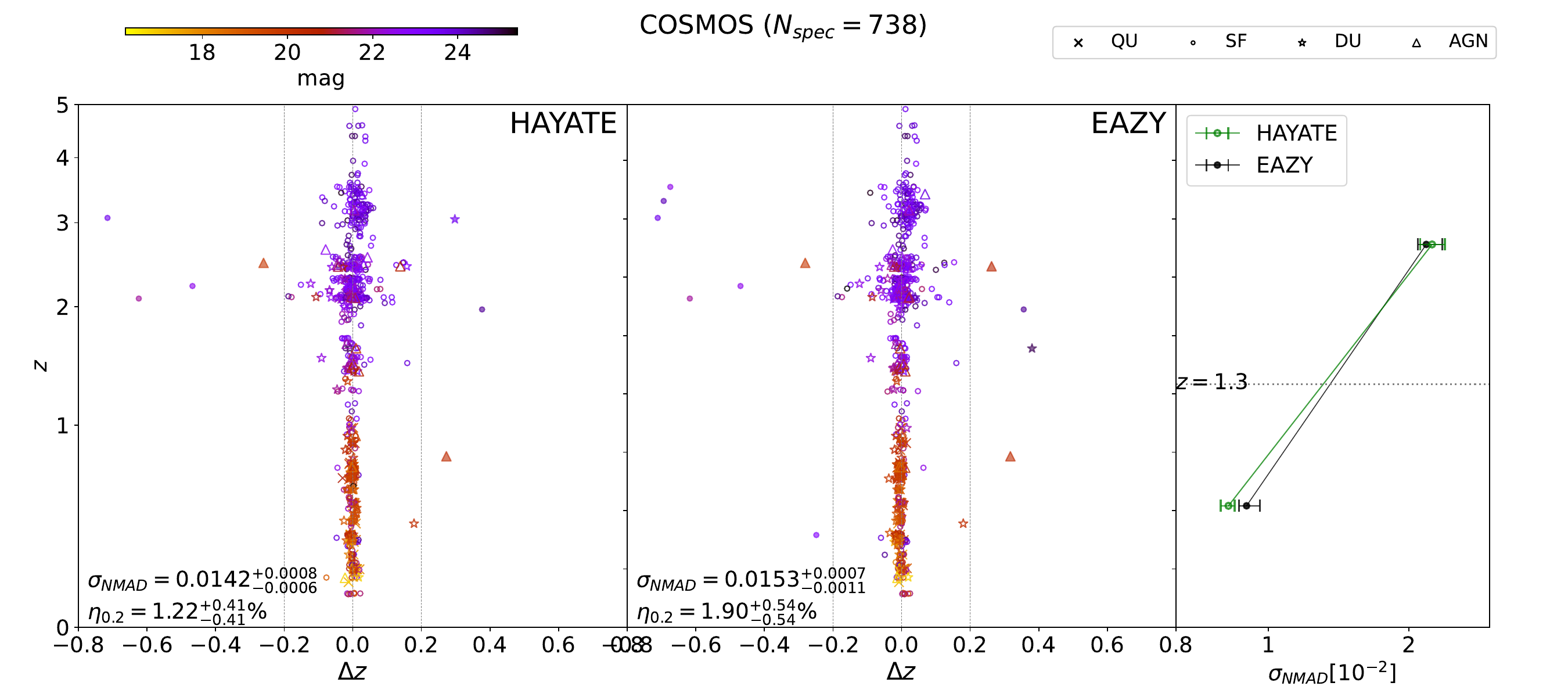}
\par\end{centering}
\begin{centering}
\includegraphics[width=1\textwidth]{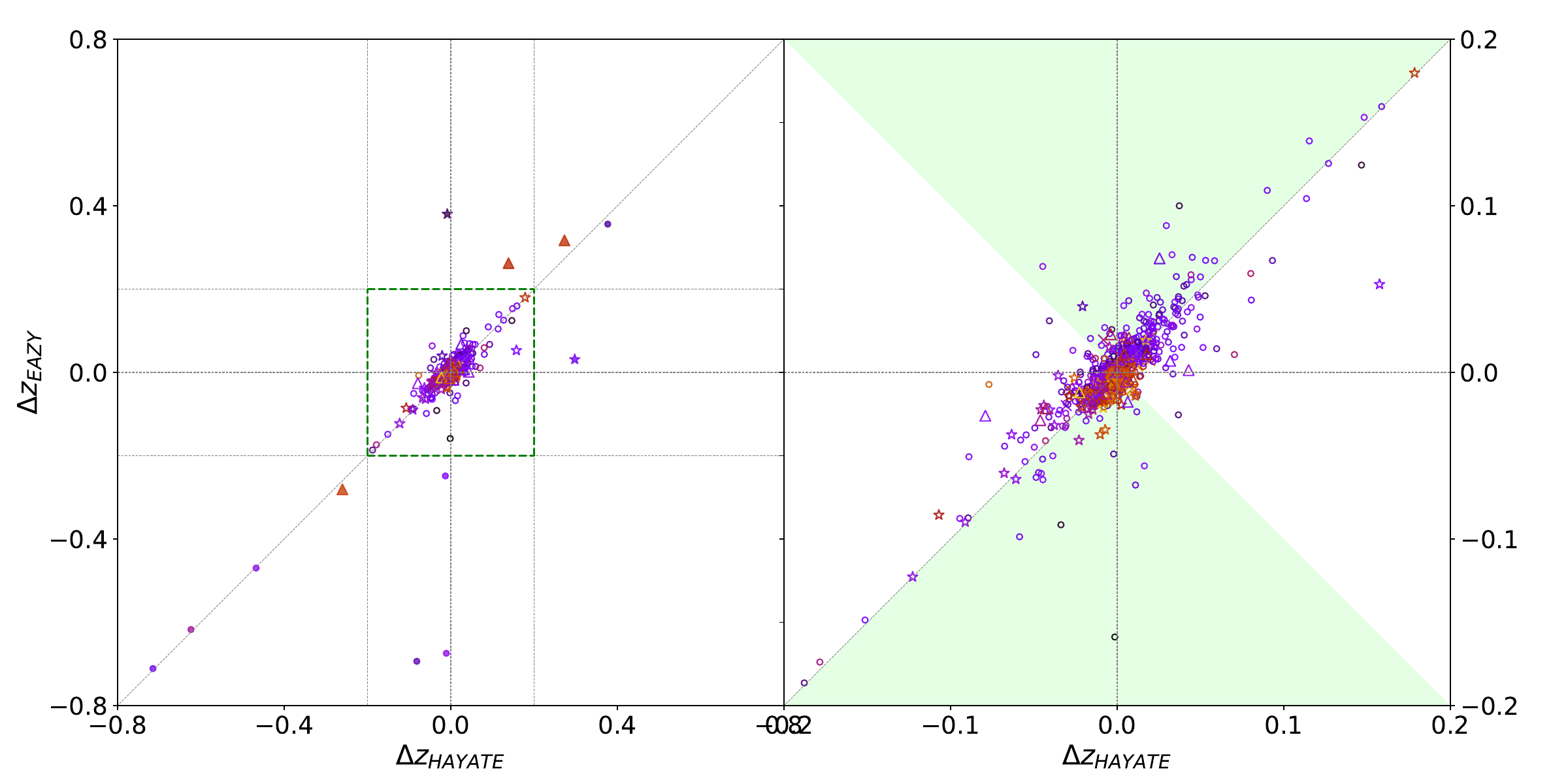}
\par\end{centering}
\caption{The same figure as Fig. \ref{fig:result_cdfs} but for COSMOS. \label{fig:result_cosmos}}
\end{figure*}

\begin{figure*}
\begin{centering}
\includegraphics[width=1\textwidth]{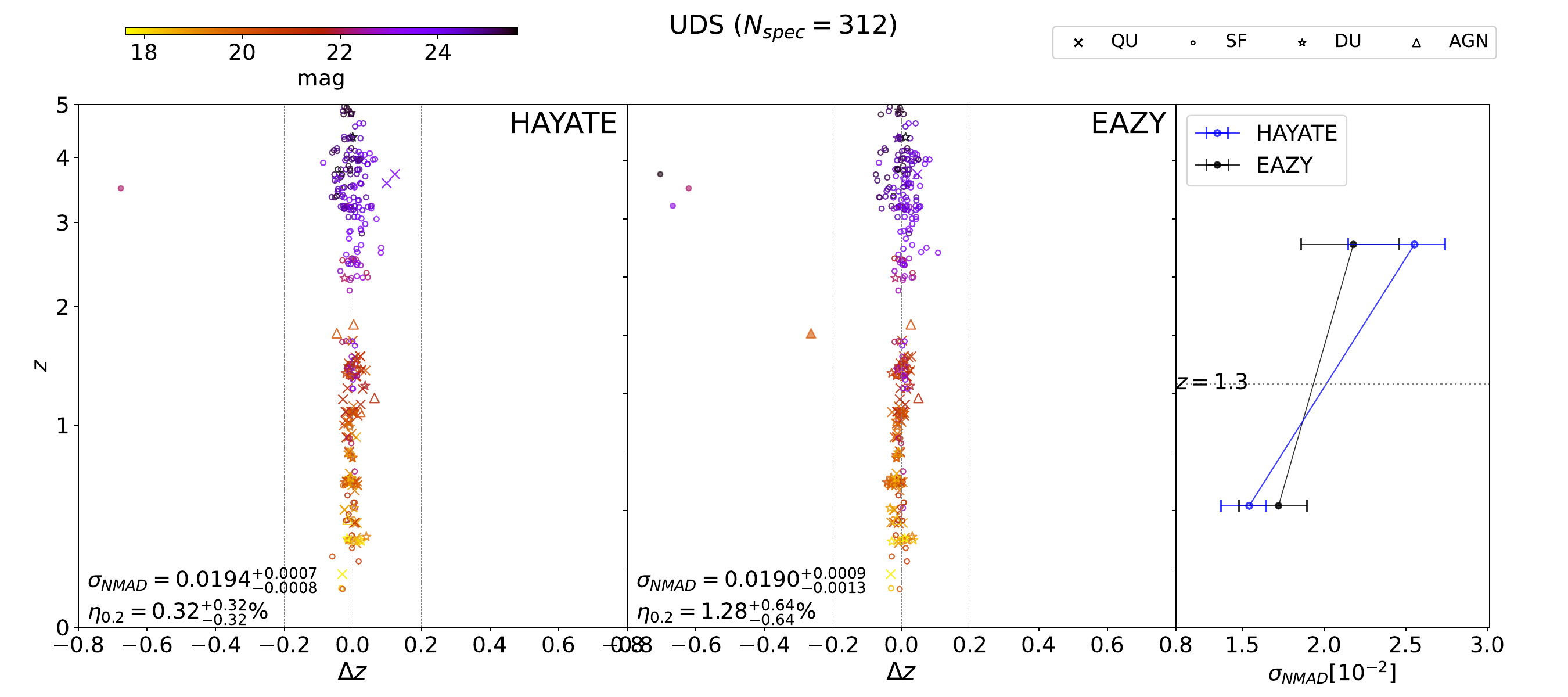}
\par\end{centering}
\begin{centering}
\includegraphics[width=1\textwidth]{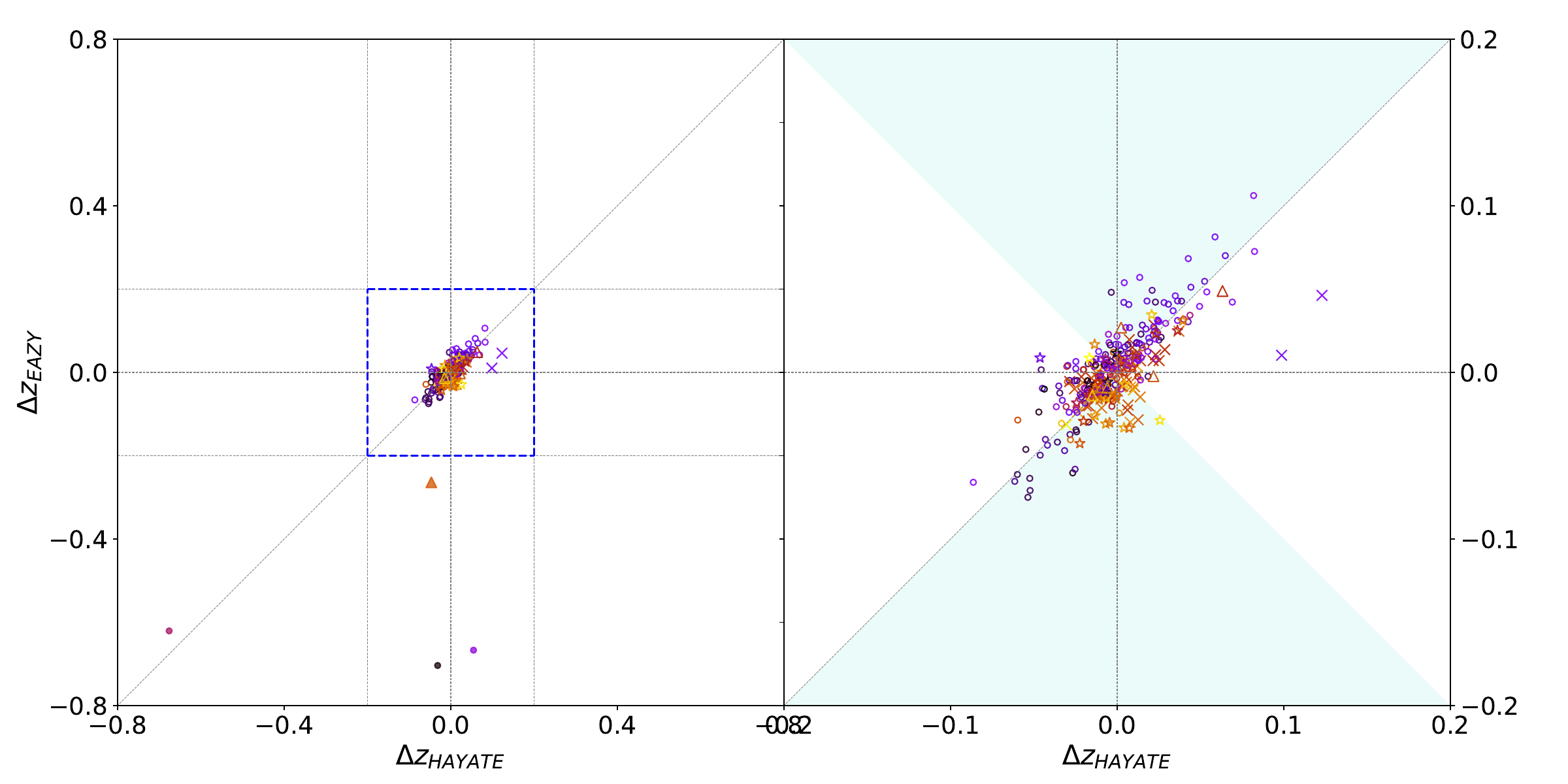}
\par\end{centering}
\caption{The same figure as Fig. \ref{fig:result_cdfs} but for UDS. \label{fig:result_uds}}
\end{figure*}

\section{Summary of photo-z FCNNs in the literature}
\label{sec:Summary_of_photoz_FCNNs}

\begin{table*}
\begin{threeparttable}
\begin{centering}
\caption{Summary of major FCNN models developed in the previous photo-z studies which were trained with spec-z information.}
\label{tab:major_fcnns}
\par\end{centering}
\begin{centering}
\begin{tabular}{ccccccc}
\hline 
\multirow{2}{*}{Reference} & \multirow{2}{*}{Architecture\tnote{(a)}} & \multicolumn{3}{c}{Target sample\tnote{(b)}} & \multicolumn{2}{c}{Photometric information\tnote{(c)}}\tabularnewline
 & & Spec-z data & Sample size & Redshift range & Survey & Filter band\tabularnewline
\hline 
1 & $\{N_{\textrm{input}}=5:$ & SDSS-EDR & $\sim7,000$ & $z_{\textrm{spec}}<0.5$ & SDSS-EDR & $ugriz$\tabularnewline
 & $6\times3:1\}$ &  &  &  &  & \tabularnewline
2 & $\{N_{\textrm{input}}=5:$ & SDSS-EDR & $\sim10,000$ & $z_{med}=0.104$ & SDSS-EDR & $ugriz$\tabularnewline
 & $10:10:1\}$ &  &  &  &  & \tabularnewline
3 & $\{N_{\textrm{input}}=7:$ & AGES & $5,052$ & $z_{\textrm{spec}}\lesssim1.0$ & NDWFS-DR3 & $B_{W}RI$\tabularnewline
 & $10\times3:1\}$ &  &  & $z_{AGN}\lesssim3.0$ & IRAC Shallow Survey & $[3.6][4.5][5.8][8.0]$\tabularnewline
4 & $\{N_{\textrm{input}}=43:$ & SDSS-DR9 & $347,342$ & $z_{\textrm{spec}}\lesssim1.0$ & SDSS-DR9 & $ugriz$\tabularnewline
 & $2N_{\textrm{input}}+1:$ &  &  &  & GALEX & $nuv,fuv$\tabularnewline
 & $N_{\textrm{input}}-1:1\}$ &  &  &  & UKIDSS & $YJHK$\tabularnewline
 &  &  &  &  & WISE & $W1,W2,W3,W4$\tabularnewline
5 & $\{N_{\textrm{input}}=5:$ & VVDS & $22,072$ & $z_{\textrm{spec}}\lesssim1.4$ & CFHTLenS & $u^{*}g'r'i'z'$\tabularnewline
 & $12:12:\mathrm{PDF}\}$\tnote{(d)} & VVDS-F22 &  &  &  & \tabularnewline
 &  & DEEP-2 &  &  &  & \tabularnewline
 &  & VIPERS &  &  &  & \tabularnewline
6 & $\{N_{\textrm{input}}=5:$ & SDSS-DR10 & $\sim180,000$ & $z_{\textrm{spec}}\lesssim0.8$ & SDSS-DR10 & $ugriz$\tabularnewline
 & $N_{\textrm{input}}+1:$ &  &  &  &  & \tabularnewline
 & $N_{\textrm{input}}+9:$ &  &  &  &  & \tabularnewline
 & $N_{\textrm{input}}+4:1\}$ &  &  &  &  & \tabularnewline
7 & $\{N_{\textrm{input}}=74:$ & SDSS-DR15 & $159,074$ & $z_{\textrm{spec}}\lesssim2.0$ & SDSS-DR15 & $ugriz$\tabularnewline
 & $528:\mathrm{PDF}\}$ &  &  & $z_{QSO}\lesssim7.0$ & WISE & $W1W2$\tabularnewline
8 & $\{N_{\textrm{input}}=4:$ & 15 spec-z & $\sim150,000$ & $z_{\textrm{spec}}\lesssim1.0$ & Pan-STARRS1-DR1 & $grizy$\tabularnewline
 & $128:256:$ & catalogues &  &  &  & \tabularnewline
 & $512:1024:$ & (e.g., SDSS-DR15 &  &  &  & \tabularnewline
 & $512:256:$ & LAMOST-DR5 &  &  &  & \tabularnewline
 & $128:32:\mathrm{PDF\}}$ & 6dFGS &  &  &  & \tabularnewline
 &  & PRIMUS) &  &  &  & \tabularnewline
\hline 
\end{tabular}
\par\end{centering}
\begin{tablenotes}
\item []\textbf{References.} (1) \cite{Firth+03}; (2) \citet[ANNz1]{Collister_Lahav04}; (3) \cite{Brodwin+06}; (4) \citet[MLPQNA]{Brescia+13,Brescia+14}; (5) \cite{Rau+15}; (6) \citet[ANNz2]{Sadeh+16}; (7) \cite{Ansari+21}; (8) \cite{Lee_Shin21};
\item []\textbf{Notes.}
\item[a] Each architecture is denoted by $\{N_{\textrm{input}}:N_1:N_{\textrm{hidden}}\times n:...:N_{\textrm{output}}\}$, where $N_{\textrm{input}}$ and $N_{\textrm{output}}$ are the numbers of input and output nodes respectively, while the first hidden layer contains $N_1$ neurons, followed by a sequence of $n$ hidden layers with $N_{\textrm{hidden}}$ neurons each, which is represented by $N_{\textrm{hidden}}\times n$.
\item[b] Spec-z sample used for evaluating the performance of a photo-z network. All the models can provide each target object with a photo-z point estimate, either as immediate output or from an output PDF of redshift. Spec-z data from different surveys can be incorporated into a collection of multiple catalogues depending on the individual work.
\item[c] Photometric data used as input. The target spec-z sample is basically limited to objects that have photometric data for all the filter bands presented in the right-most column.
\item[d] $N_{\textrm{output}}=1$ implies that the model predicts a photo-z point estimate for each object, while $N_{\textrm{output}}=\textrm{PDF}$ it produces a PDF of redshift as immediate output.
\end{tablenotes}
\end{threeparttable}
\end{table*}

\begin{table*}
\begin{threeparttable}
\begin{centering}
\caption{Summary of photo-z FCNN models which were trained with simulations.}
\label{tab:fcnns_w_simulations}
\par\end{centering}
\begin{centering}
\begin{tabular}{ccccccc}
\hline 
\multirow{2}{*}{Reference} & \multirow{2}{*}{Architecture} & \multicolumn{3}{c}{Target sample} & \multicolumn{2}{c}{Photometric information}\tabularnewline
 &  & Spec-z data & Sample size & Redshift range & Survey & Filter band\tabularnewline
\hline 
1 & $\{N_{\textrm{input}}=7:$ & SDSS-DR1 & $88,108$ & $z_{\textrm{spec}}\lesssim0.4$ & SDSS-DR1 & $ugriz$\tabularnewline
 & $12:10:1\}$ &  &  &  &  & \tabularnewline
2 & $\{N_{\textrm{input}}=46:$ & zCOSMOS-DR3 & $8,566$ & $z_{\textrm{spec}}\lesssim1.2$ & PAUS & 40 narrow bands\tabularnewline
 & $600:400:$ &  &  &  &  & $(4500-8500\mathrm{\text{\AA}})$\tabularnewline
 & $250\times13:\mathrm{PDF}\}$ &  &  &  & CFHTLenS & $u^{*}$\tabularnewline
 &  &  &  &  & COSMOS-20 & $BVri^{+}z^{++}$\tabularnewline
3 & $\{N_{\textrm{input}}=5:$ & SDSS-DR15 & $1,965,800$ & $z_{\textrm{spec}}\lesssim1.0$ & SDSS-DR15 & $ugriz$\tabularnewline
 & $512:1024:$ & VIPERS &  &  & CFHTLS & $ugriz$\tabularnewline
 & $2048:1024:$ & DEEP-2 &  &  &  & \tabularnewline
 & $512:256:128$ &  &  &  &  & \tabularnewline
 & $64:32:\mathrm{PDF}\}$ &  &  &  &  & \tabularnewline
4 & $\{N_{\textrm{input}}=20:$ & Simulation\tnote{(a)} & $\sim10,000$ & $z_{\textrm{sim}}\lesssim4.0$ & CSST & $NUV,ugrizy$\tabularnewline
 & $2N_{\textrm{input}}\times2:1\}$ &  &  &  &  & \tabularnewline
 & $\{N_{\textrm{input}}=20:$ &  &  &  &  & \tabularnewline
 & $2N_{\textrm{input}}\times6:1\}$ &  &  &  &  & \tabularnewline
This & $\{N_{\textrm{input}}=76:$ & S16 & $1,273$ & $z_{\textrm{spec}}<5.0$ & ZFOURGE & $J_{1}J_{2}J_{3}H_{s}H_{l}K_{s}$\tabularnewline
work & $500\times3:\mathrm{PDF}\}$ & (e.g. FORS2 &  &  & HUGS & $KsHI$\tabularnewline
(CDFS) &  & K20 &  &  & TENIS & $tenisK$\tabularnewline
 &  & VVDS &  &  & VIMOS & $U,R$\tabularnewline
 &  & CXO &  &  & ACS & $BVIZ$\tabularnewline
 &  & IMAGES &  &  & ESO DPS & $U_{38}VR_{c}$\tabularnewline
 &  & VIMOS) &  &  & 3D-HST & $F140W,F814W$\tabularnewline
 &  & Update on $z_{\textrm{spec}}$ data &  &  & WFC3 ERS & $F098M,F125W,F160W$\tabularnewline
 &  & (MOSDEF &  &  & CANDELS & $F105M,F125W,F160W$\tabularnewline
 &  & MOSEL &  &  &  & $F606W,F814W$\tabularnewline
 &  & VANDELS) &  &  & MUSYC & $IA484,IA527,IA574$\tabularnewline
 &  &  &  &  &  & $IA598,IA624,IA651$\tabularnewline
 &  &  &  &  &  & $IA679,IA738,IA767$\tabularnewline
 &  &  &  &  &  & $IA797,IA856$\tabularnewline
 &  &  &  &  & IUDF & $[3.6][4.5]$\tabularnewline
 &  &  &  &  & GOODS & $[5.8][8.0]$\tabularnewline
\hline 
\end{tabular}
\par\end{centering}
\begin{tablenotes}
\item []\textbf{References.} (1) \cite{Vanzella+04}; (2) \citet[DEEPZ]{Eriksen+20}; (3) \citet[SYTH-Z]{Ramachandra+22}; (4) \cite{Zhou+21,Zhou+22};
\item [] \textbf{Notes.} 
\item[a] The model was tested on CSST mock data simulated in the future surveys.
\end{tablenotes}
\end{threeparttable}
\end{table*}

\bsp	
\label{lastpage}
\end{document}